\documentclass[a4paper,11pt]{article}
\pdfoutput=1 

\usepackage{jheppub} 
\usepackage{graphicx}
\usepackage{latexsym}
\usepackage{bbm}
\usepackage[mathscr]{euscript}
\usepackage{subcaption}
\usepackage{amsmath,amssymb,bm} 
\newcommand{\beq}{\begin{equation}}
\newcommand{\eeq}{\end{equation}}
\newcommand{\ba}{\begin{array}{ccc}}

\def\bea{\begin{eqnarray}}
\def\eea{\end{eqnarray}}
\usepackage[colorlinks=true]{hyperref} 
\hypersetup{
    bookmarks=true,         
    unicode=false,          
    pdftoolbar=true,        
    pdfmenubar=true,        
    pdffitwindow=false,     
    pdfstartview={FitH},    
    pdfsubject={},   
    pdfcreator={},   
    pdfproducer={}, 
    pdfkeywords={} {} {}, 
    pdfnewwindow=true,      
    colorlinks=true,       
    linkcolor=magenta, 
    citecolor=blue,        
    filecolor=magenta,      
    urlcolor=blue           
}

\renewcommand{\approx}{\simeq}

\title{Scaling limits of complex Sachdev-Ye-Kitaev models\\ and holographic geometry}

\author[a]{Elena Gubankova}
\author[a,b]{Subir Sachdev}
\author[c]{Grigory  Tarnopolsky}
\affiliation[a]{Department of Physics, Harvard University, Cambridge, MA 02138,USA}
\affiliation[b]{Center for Computational Quantum Physics, Flatiron Institute, 162 5th Avenue, New York, NY 10010, USA}
\affiliation[c]{Department of Physics, Carnegie Mellon University, Pittsburgh, PA 15213, USA}

\emailAdd{egoubankova@fas.harvard.edu}
\emailAdd{sachdev@g.harvard.edu}
\emailAdd{gtarnopo@andrew.cmu.edu}

\abstract{We compare different limits of  the Sachdev-Ye-Kitaev model of $N$ complex fermion with $p$-fermion interactions. First, we compute the fermion Green's function and free energy in the limit of large $N$ followed subsequently by the limit of large $p$. Next, we examine the `double-scaling' limit in which the large $N,p$ limits are taken at fixed $\lambda = p^2/N$. Earlier results on the latter limit are resummed for small $\lambda$, and shown to match our results for the first limit. We also describe the holographic match of our results to two-dimensional Jackiw-Teitelboim gravity with an additional $U(1)$ gauge field.}

\begin{document}
\maketitle
\flushbottom

\section{Intoduction}

The Sachdev-Ye-Kitaev (SYK) model is a quantum mechanical model (0 + 1 dimensions) of $N$ Majorana fermions $\psi_i$ with random all-to-all interactions, in groups of $p$ fermions, 
and the couplings are taken to be random. The SYK model is a highly chaotic, with a maximal quantum
Lyapunov exponent at low temperatures. 
It can be
studied analytically in the large $N$ limit. The Hamiltonian in Majorana case is given by
\begin{align}
H =i^{p/2}\sum_{1\leq i_1<  \cdots < i_p\leq N } J_{i_1\cdots i_p} \psi_{i_1}\cdots \psi_{i_p}    \,,
\end{align}
where $p$ is a parameter in the model which sets the length of the all-to-all interactions in the
Hamiltonian, and the $J_{i_1\cdots i_p}$ are the random couplings.
There are two cases where the SYK model can be solved analytically. First, the SYK model develops an approximate conformal
symmetry in the infrared, {\it i.e.\/} at small temperatures (low energies) and strong coupling $\beta J\gg1$ with $\beta$ being an inverse temperature \cite{SYK_Review,SYK_Review2,CGPS}. Second,  
the model simplifies considerably in the limit of large $p$. In this case results are valid beyond the CFT regime and at all energy scales.

In the large $p$ limit two alternative techniques have been developed. In one approach $N$ and $p$ are treated as independent parameters where first $N$ is taken to infinity resulting in the set of Schwinger-Dyson equations. 
Then as $p$ goes to infinity, SYK gives the Liouville's theory for the two-point function \cite{Large_q}. In the other approach, known as double-scaled SYK, $p$ scales with $N$, 
while combination 
\begin{equation}
\lambda =  \frac{p^2}{N}    
\end{equation} 
is kept fixed providing an independent parameter $\lambda$. In this limit the model can be solved exactly at all energy scales using combinatorial tools. 
Computation of the correlation functions in the double-scaled SYK reduces to a counting problem of chord diagrams. This counting problem and summation of chord diagrams can be solved 
by introducing the transfer matrix integrating over the energies \cite{Double_scaled_SYK,DSSYK2,DSSYK_Review}.
The double scaled SYK is controlled by a quantum group symmetry which replaces the conformal symmetry at all energy scales. 
Due to the quantum group symmetry the double-scaled SYK provides an insight in holographic dual description
and connection to quantum gravity. Finite $\lambda$ correlation functions of the double-scaled SYK have been obtained in the form of multivariable integrals over energies that are hard to track analytically,  
for the Majorana fermions \cite{Double_scaled_SYK,DSSYK2} and in the complex case \cite{Double_scaled_complex_SYK}. 

In the limit of vanishing parameter $\lambda$ the integrands of the double-scaled SYK become sharply peaked and dominated by a saddle point. 
At low energies and in the limit $\lambda\to 0$ the model is described by the Schwarzian theory. There is a concept that two methods: direct large $p$ calculations and the double-scaled SYK in the limit $\lambda\to 0$
should produce the same correlation functions. Indeed this agreement was verified for the Majorana fermions \cite{Double_scaled_SYK_small_lambda,Double_scaled_SYK_small_lambda2,Double_scaled_SYK_small_lambda3}.
In this paper we verify this concept for the complex SYK, {\it i.e.\/} for the SYK fermions at nonzero charge density.

SYK models also have a holographic connection to black holes with an AdS$_2$ sector in their near horizon geometries, which was initially noted semiclassically \cite{SS10}, but holds also at the fully quantum level \cite{Kitaev_Suh,Large_q,Large_q_Schwarzian}. Here we will further explore the connection for the complex SYK model, which requires a U(1) gauge field in the bulk \cite{Complex_SYK,Sachdev:2019bjn,Complex_SYK-Large_q,Complex_SYK_Density_of_states}.

\subsection{Organization of the paper}
This paper is organized as follows. In Section \ref{section_largepSYK} we calculate  the two-point Green's function and thermodynamic grand potential of the complex SYK model in the large $p$ limit. In Section \ref{section_DSSYK} we calculate the partition and two-point correlation functions in the double-scaled complex SYK model in the limit $\lambda\to 0$, and reproduce the  large $p$ results of Section \ref{section_largepSYK}. 
In Section \ref{section_metric} we find the bulk geometry induced in the small $\lambda$ limit; for the complex SYK model this requires a $U(1)$ gauge field that is not present in the Majorana case. We also review relevant aspects of the complex SYK model in the Appendices.

\section{Large \texorpdfstring{$p$}{p} solution for the complex SYK model}
\label{section_largepSYK}

The Hamiltonian for the complex SYK model is given by
\begin{align}
H =\sum_{\substack{  1\leq i_1<  \ldots < i_{p/2}\leq N \\ 1\leq j_1<  \ldots < j_{p/2}\leq N}}
J^{i_1\cdots i_{p/2}}_{j_1\cdots j_{p/2}} \,\bar{\psi}_{i_1}\cdots \bar{\psi}_{i_{p/2}} \psi^{j_1}\cdots \psi^{j_{p/2}}  - \mu \sum_{j} \bar{\psi}_j \psi^{j}\,,   
\label{complex_Hamiltonian}
\end{align}
where $\mu$ is the fermion chemical potential and the complex fermions $\bar{\psi}_i$ and $\psi^i$ satisfy 
$\{\psi^i,\bar{\psi}_j\}=\delta^i_j$,
$\{\psi^i,\psi^j\}=\{\bar{\psi}_i,\bar{\psi}_j\}=0$ with
$i,j=1,2,\ldots,  N$. The total number of (anti)fermions participating in all-to-all interactions at a given instant is $p$, which sets the ``length'' of the SYK Hamiltonian operator. 
The couplings $J^{i_1\cdots i_{p/2}}_{j_1\cdots j_{p/2}}$ are Gaussian complex random variables with zero mean satisfying
$(J^{i_1\cdots i_{p/2}}_{j_1\cdots j_{p/2}})^{\ast}=J_{i_1\cdots i_{p/2}}^{j_1\cdots j_{p/2}}$, ensuring the hermiticity of the Hamiltonian. Their variance is \cite{Complex_SYK_Density_of_states}
\begin{align}
\left\langle J^{i_1\cdots i_{p/2}}_{j_1\cdots j_{p/2}}
J_{i_1\cdots i_{p/2}}^{j_1\cdots j_{p/2}}\right\rangle_J
=J^2\frac{(p/2)!(p/2-1)!}{N^{p-1}}\,. \label{JJvar}
\end{align}
We define the conserved $U(1)$ charge density operator $\hat{Q}$ by
\begin{align}
\hat{Q} \equiv  \frac{1}{N} \sum_{j} \bar{\psi}_{j}\psi^{j} -\frac{1}{2}\,,
\end{align}
and it is related to the asymmetry of the two point Green's function at finite temperature 
\begin{align}
G(\tau)=\frac{1}{N}\sum_{j}\langle \textrm{T} \psi^{j}(\tau) \bar{\psi}_{j}(0)\rangle, \quad G(0^{+}) = \frac{1}{2}-Q, \quad G(\beta^{-})=\frac{1}{2}+Q\,, \label{GandQ}
\end{align}
where $Q \equiv \langle \hat{Q} \rangle$ and $\beta=1/T$ is the inverse temperature. 
The parameter $J$ in (\ref{JJvar}) is a normalization constant for the disorder. To get a standard form of the Liouville equation for the Green's function in the large $p$ limit \cite{Large_q} we replace $J\rightarrow \mathcal{J}$ 
\begin{align}
\mathcal{J}^2 \equiv \frac{pJ^2}{2\big(2\cosh\frac{\beta\mu}{2}\big)^{p-2}}\,.
\end{align}
Using the Stirling's approximation the variance has an asymptotic behavior at large $p$ 
\begin{align}
\left\langle J^{i_1\cdots i_{p/2}}_{j_1\cdots j_{p/2}}
J_{i_1\cdots i_{p/2}}^{j_1\cdots j_{p/2}}\right\rangle_J
=\mathcal{J}^2\frac{\pi \, p^{p-1}e^{-p}}{N^{p-1}}\left(\cosh\frac{\beta\mu}{2}\right)^{p-2}.
\label{variance2}
\end{align}
In this section we use the fact that the SYK model simplifies considerably in the limit of large $p$. Considering the Schwinger-Dyson equations valid in the limit $N\to \infty$ we then take the limit $p\to\infty$
and use an ansatz for the two-point Green's function 
suggested in Refs.~\cite{Large_q} and \cite{Large_q_Liouville}. We calculate the Green's function and thermodynamic grand potential 
by the method first used for the Majorana fermions \cite{Large_q}. As we will show, our calculations are valid in the range of chemical potentials
\begin{align}
Q_0^2 \leq \frac{1}{2e} \frac{1}{\beta \mathcal{J}}\,,
\label{mu_range}
\end{align}
where  $e\approx 2.718$ is the Euler's number and 
\begin{align}
    Q_0=\frac{1}{2}\tanh\Big(\frac{\beta\mu}{2}\Big)
\end{align}
is the charge density for free fermions.
In the  regime $\beta \mathcal{J}\gg 1$
Eq.~(\ref{mu_range}) corresponds to small values of chemical potential
\begin{align}
(\beta\mu)^2 \leq  \frac{8}{e} \frac{1}{\beta \mathcal{J}}\,.
\label{}
\end{align}

\subsection{Complex SYK  Green's function in the large $p$ limit}
\label{sec:largepG}
The large $N$ Schwinger-Dyson equations for the complex SYK model (\ref{complex_Hamiltonian}) can be written in the form  (see Appendix \ref{cSYKEffAct} for a review)
\begin{align}
G(i\omega_{n})= \frac{1}{-i\omega_{n}-\mu-\Sigma(i\omega_{n})}, \quad \Sigma(\tau) = J^{2}G(\tau)^{\frac{p}{2}} G(\beta-\tau)^{\frac{p}{2}-1}\,,
\label{SD0}
\end{align}
where $\omega_{n}=2\pi/\beta(n+1/2)$ are the Matsubara frequencies. These equations imply the following boundary conditions for the Green's function:
\begin{align}
&G(0^{+})+G(\beta^{-})=1\,, \quad \partial_{\tau}G(0^{+})+\partial_{\tau}G(\beta^{-})  = \mu\,. \label{BC0}
\end{align}
The plausible ansatz for the large $p$ solution reads \cite{Large_q}
\begin{align}
G(\tau) =G_{0}(\tau) \Big(1+\frac{1}{p}g(\tau) + O(1/p^2)\Big)\,, 
\label{largeqanz}
\end{align}
where $G_{0}(\tau)$ is the free fermion Green's function and is given by
\begin{align}
G_{0}(\tau) =  
\begin{cases}
\frac{e^{\mu\tau}}{1+e^{ \beta\mu }},& 0< \tau < \beta  \\
 - \frac{e^{\mu\tau}}{1+e^{- \beta \mu }},& -\beta < \tau < 0 
 \end{cases} \,.  \label{G0function}
\end{align}

The main new result of this section is Eq.~(\ref{g_solution}) for $g(\tau)$.  We now outline its derivation. 
Using the ansatz Eq.~(\ref{largeqanz}) one can compute $\Sigma(\tau)$ in the large $p$ limit using Eq.~(\ref{SD0})
\begin{align}
\Sigma (\tau)  =  \frac{J^{2}}{(2 \cosh \frac{\beta\mu }{2})^{p-2}} G_{0}(\tau) e^{\frac{1}{2}(g(\tau)+g(\beta-\tau))} +O(1/p)\,,
\label{Sigma}
\end{align}
where we used that
\begin{equation}
G_0(\tau)G_0(\beta-\tau)=-\frac{1}{(2\cosh\frac{\beta\mu}{2})^2}\,.
\label{G0G0}
\end{equation}
Using the Green’s-function ansatz (\ref{largeqanz}), the first Schwinger–Dyson equation in (\ref{SD0}) gives
\begin{align}
\frac{1}{G(i\omega_{n})} = \frac{1}{\frac{1}{-i \omega_{n}-\mu}+\frac{1}{p}[G_{0}\times g](\omega)} =-i \omega_{n}-\mu - \Sigma(i \omega_{n})\,,
\end{align}
where we used Eq.~(\ref{G0function}) for the free fermion Green's function.
Expanding the denominator in $1/p$ series we find 
\begin{align}
-\frac{1}{p}(-i \omega_{n}-\mu)^{2}[G_{0}\times g](\omega) = - \Sigma(i \omega_{n})\,.
\end{align}
Returning to the imaginary-time representation, we obtain the differential equation
\begin{align}
(\partial_{\tau}-\mu)^{2}(G_{0}(\tau) g(\tau))= p \Sigma(\tau)\,.
\end{align}
Using the explicit form of the free fermion Green’s function $G_{0}(\tau)$, this becomes
\begin{align}
G_{0}(\tau) \partial^{2}_{\tau} g(\tau)=  \frac{pJ^{2}}{(2 \cosh \frac{\mu \beta}{2})^{p-2}} G_{0}(\tau) e^{\frac{1}{2}(g(\tau)+g(\beta-\tau))}\,.
\end{align}
Therefore, for $0<\tau<\beta$ the function $g(\tau)$ satisfies the Liouville-type equation:
\begin{align}
 \partial^{2}_{\tau} g(\tau)= 2 \mathcal{J}^{2} e^{\frac{1}{2}(g(\tau)+g(\beta-\tau))}\,, \quad  \mathcal{J}^{2} \equiv  \frac{pJ^{2}}{2\big(2 \cosh \frac{ \beta \mu}{2}\big)^{p-2}}\,.
 \label{LE}
\end{align}

Since the right hand side of this differential equation is symmetric under $\tau \to \beta -\tau$, it follows that
\begin{align}
 \partial^{2}_{\tau} g(\tau)=  \partial^{2}_{\tau} g(\beta -\tau) \,.
 \label{sym}
\end{align}
Therefore we can rewrite Eq.~(\ref{LE}) in the symmetric form
\begin{align}
 \partial^{2}_{\tau} \left[\frac{1}{2}(g(\tau)+g(\beta-\tau))\right]=  2\mathcal{J}^{2} e^{\frac{1}{2}(g(\tau)+g(\beta-\tau))}.
 \label{LE2}
\end{align}
We can introduce symmetric and antisymmetric Green's function fluctuations with respect to permutation $\tau\leftrightarrow (\beta-\tau)$
\begin{align}
g_s(\tau)=\frac{1}{2}\left(g(\tau)+g(\beta-\tau)\right),\quad g_a(\tau)=\frac{1}{2}\left(g(\tau)-g(\beta-\tau)\right).
 \label{}
\end{align} 
From Eqs.~(\ref{sym},\ref{LE2}) the equations of motion are
\begin{align}
\partial^{2}_{\tau} g_s(\tau)=  2\mathcal{J}^{2} e^{g_s(\tau)},\quad \partial^{2}_{\tau} g_a(\tau)=0, 
\label{eqs_of_motion}
\end{align} 
with corresponding boundary conditions.\footnote{As we submitted our paper we became aware of paper \cite{Arundine:2025mcu} where similar equations appear. Our derivation was carried out independently, before that work became publicly available.} For the Majorana fermions symmetric part vanishes at the boundary \cite{Large_q}, while in the complex case it is nonzero $g_s(0^+)= g_s(\beta^-)\neq 0$.
It follows from Eq.~(\ref{sym}) $ \partial_{\tau} g(\tau)=\partial_{\tau} g(\beta -\tau)+C_{1}$ and taking $\tau =\beta/2$ we get that $C_{1}=2 \partial_{\tau} g(\beta/2)$. 
Integrating one more time  we obtain that $g(\tau)=g(\beta-\tau)+C_{1} \tau +C_{2}$ and again taking $\tau =\beta/2$ we get that $C_{2}=-\beta C_{1}/2 = -\beta \partial_{\tau} g(\beta/2) $. So finally we find that 
\begin{align}
g(\tau)=g(\beta-\tau)+ ( 2\tau- \beta)\partial_{\tau}g(\beta/2)\,. \label{relg1}
\end{align}

From the boundary conditions Eq.~(\ref{BC0}) we find using Eq.~(\ref{GandQ})
\begin{eqnarray}
&&\frac{1}{2}\left[g(0^{+})+g(\beta^{-})\right]-Q_0\left[g(0^{+})-g(\beta^{-})\right]=0,\\
&& \frac{1}{2}\left[\partial_{\tau} g(0^{+})+\partial_{\tau}g(\beta^{-})\right]-Q_0\left[\partial_{\tau}g(0^{+})-\partial_{\tau}g(\beta^{-})\right]=0
\end{eqnarray}
that gives the relations 
\begin{align}
g(\beta^{-})= -e^{-\beta \mu}g(0^{+}), \quad \partial_{\tau}g(\beta^{-})= -e^{-\beta \mu} \partial_{\tau}g (0^{+})\,. \label{grel2}
\end{align}
Thus from (\ref{relg1}) and (\ref{grel2}) we find 
\begin{align}
g(0^{+})= -\beta\Big(Q_{0}+\frac{1}{2}\Big)  \partial_{\tau}g(\beta/2), \quad 
g(\beta^{-})= -\beta  \Big(Q_{0}-\frac{1}{2}\Big) \partial_{\tau}g(\beta/2),
\label{grel3}
\end{align}
therefore
\begin{align}
 \frac{1}{2}(g(0^{+})+g(\beta^{-}))=-\beta Q_{0}  \partial_{\tau}g(\beta/2) ,
 \label{boundary}
 \end{align} 
 and
\begin{align}
\partial_{\tau}g(0^{+}) =\frac{2}{1-e^{-\beta\mu }} \partial_{\tau}g(\beta/2), \quad
\partial_{\tau}g(\beta^{-}) =\frac{2}{1-e^{\beta\mu }}\textbf{}g(\beta/2).
\label{grel33}
\end{align}
Using these relations we have the following equations with the boundary conditions for the symmetric part
\begin{eqnarray}
\left\{\begin{array}{@{}ll@{}}
\partial^{2}_{\tau} g_s(\tau) =  2\mathcal{J}^{2} e^{g_s(\tau)}  \label{equation_gs}\\
g_s(0^{+}) = g_s(\beta^{-}) = -\beta Q_0 \partial_{\tau}g(\beta/2)
\end{array}\right.
\label{symmetric}
\end{eqnarray}
and the antisymmetric part
\begin{eqnarray}
\left\{\begin{array}{@{}ll@{}}
\partial^{2}_{\tau} g_a(\tau) =  0    \label{equation_ga}\\
g_a(\beta/2) = 0,\,\, \partial_{\tau}g_a(\beta/2)=\partial_{\tau}g(\beta/2)
\end{array}\right.
\label{antisymmetric}
\end{eqnarray}
Now we are ready to solve the equation (\ref{LE2}). Define a new function 
\begin{align}
h(\tau) \equiv \frac{1}{2}(g(\tau)+g(\beta-\tau))+\beta Q_{0}   \partial_{\tau}g(\beta/2) \,. \label{defh}
\end{align}
This function satisfies the Liouville equation 
\begin{align}
\begin{cases}
 \partial^{2}_{\tau} h(\tau) =2 \mathcal{J}^{2} e^{-\beta Q_{0} \partial_{\tau}g(\beta/2)  } e^{h(\tau)}\\
 h(0^{+}) = h(\beta^{-})=0 
 \end{cases}\,, \label{Liouville}
 \end{align}
which has zero boundary conditions due to Eq.~(\ref{boundary}). Using Eq.~(\ref{relg1})  and the definition of $h(\tau)$ in (\ref{defh}) we can eliminate $g(\beta-\tau)$ and find that 
\begin{align}
g(\tau)=\Big(\tau-\beta\big(Q_0+\frac{1}{2}\big)\Big)\partial_{\tau} g(\beta/2)+h(\tau),  \label{gtoh}
\end{align}
with $h(\tau)$  is a solution of Eq.~(\ref{Liouville}), and $\partial_{\tau}g(\beta/2)$ can be expressed through $h(\tau)$ when it is known.

We know that the equation (\ref{Liouville}) has the solution \cite{Large_q}
\begin{align}
h(\tau) = \log \bigg(\frac{\cos^{2}\big(\frac{\pi v}{2}\big)}{\cos^{2}\big(\frac{\pi v}{2}-\frac{\pi v \tau}{\beta}\big)}\bigg), \quad \frac{\pi v}{\cos \big(\frac{\pi v}{2}\big)} = \beta  \mathcal{J}  e^{-\frac{1}{2}\beta Q_{0} \partial_{\tau}g(\beta/2) }\,,
\label{h}
 \end{align}
therefore using Eq.~(\ref{gtoh}), we have 
 \begin{align}
g(\tau) =\Big(\tau - \beta(Q_{0}+\frac{1}{2})\Big) \partial_{\tau}g(\beta/2) +\log \bigg(\frac{\cos^{2}\big(\frac{\pi v}{2}\big)}{\cos^{2}\big(\frac{\pi v}{2}-\frac{\pi v \tau}{\beta}\big)}\bigg) \,.
 \end{align}
Differentiating this equation by $\tau$ we obtain 
\begin{align}
\partial_{\tau}g(\tau) = \partial_{\tau}g(\beta/2) - \frac{2\pi v}{\beta} \tan\Big(\frac{\pi v}{2}-\frac{\pi v \tau}{\beta}\Big)
 \end{align}
Thus at $\tau = 0^{+}$ we find that $ \partial_{\tau}g(0^{+}) =\partial_{\tau}g(\beta/2) - \frac{2\pi v}{\beta} \tan\frac{\pi v}{2} $, and using (\ref{grel33}) we finally find $\partial_{\tau}g(\beta/2)$
\begin{align}
\partial_{\tau}g(\beta/2) = - Q_{0}\frac{4\pi v}{\beta} \tan \frac{\pi v}{2}\,.
\label{dotg}
 \end{align}
 Finally, the whole solution reads
\begin{align}
g(\tau) =  - \frac{4\pi v}{\beta} \tan \big(\frac{\pi v}{2}\big)Q_{0}\Big(\tau - \beta(Q_{0}+\frac{1}{2})\Big) +\log \bigg(\frac{\cos^{2}\frac{\pi v}{2}}{\cos^{2}\big(\frac{\pi v}{2}-\frac{\pi v \tau}{\beta}\big)}\bigg) \,,
\label{g_solution}
 \end{align}
where $v$ is found from the transcendental equation 
\begin{align}
\frac{\pi v}{\cos\big( \frac{\pi v}{2}\big)}e^{- 2\pi v \tan( \frac{\pi v}{2}) Q_{0}^{2} } = \beta  \mathcal{J}  , \quad \mathcal{J}^{2} \equiv \frac{pJ^{2}}{2\left(2 \cosh \frac{\beta\mu }{2}\right)^{p-2}}\,,
\label{betaJ_solution}
\end{align}
and we used Eqs.~(\ref{h},\ref{dotg}). Compared to the solution for Majorana fermions \cite{Large_q}, which is given only by the logarithm term in Eq.~(\ref{g_solution}), the Green's function $g(\tau)$ in the complex SYK model has extra antisymmetric linear in time term $g_a(\tau)\sim Q_0(\tau-\frac{\beta}{2})$
and nonzero boundary value of symmetric part $g_s(0)=g_s(\beta)\sim Q_0^2\beta$, which are given by the first term in Eq.~(\ref{g_solution}) (we present a separate solution for $g_{s}$ in Appendix \ref{SymGreenSolv}). 
The antisymmetric term $g_a$ will be connected to the $U(1)$ gauge field in the holographic discussion of Section~\ref{section_metric}. 
 
We discuss different $\beta \mathcal{J}$ limits of the large $p$ solution Eq.~(\ref{g_solution}) and also its comparison with the conformal solution in the next subsection \ref{LargepandGc}. There we show that solution of Eq.~(\ref{betaJ_solution}) exists for $Q_{0}^2 \leq 1/(2 e \beta \mathcal{J})$.  We compare the large $p$ approximation with exact large $N$ complex SYK Green's function in the Appendix \ref{GexactGp}.

 
\subsection{Analysis of the large \texorpdfstring{$p$}{p} Green's function}
\label{LargepandGc}

Here we discuss weak and strong coupling limits of the large \texorpdfstring{$p$}{p} Green's function and make comparison with the conformal solution.
The large $p$ solution Eqs.~(\ref{largeqanz}, \ref{g_solution}),  in exponentiated form is
\begin{align}
G_{\textrm{large}\, p}(\tau) =G_{0}(\tau) e^{\frac{4\pi v}{p}\tan( \frac{\pi v}{2})  Q_{0}^2} \left(\frac{\cos\frac{\pi v}{2}}{\cos(\frac{\pi v}{2}-\frac{\pi v \tau}{\beta})}\right)^{2/p} e^{-\frac{4\pi v}{p}\tan (\frac{\pi v}{2})   Q_{0}(\frac{\tau}{\beta}-\frac{1}{2})}\,,
\label{largepmu}
\end{align}
where $G_{0}(\tau)$ is the free fermion Green’s function Eq.~(\ref{G0function}), $Q_{0}=\tanh(\beta \mu/2)/2$ and  $v$ is found from solving the transcendental Eq.~(\ref{betaJ_solution}).
For small $\beta \mathcal{J}$, $v$ admits the expansion 
\begin{align}
v = \frac{1}{\pi} \beta \mathcal{J} -\frac{1-8Q_{0}^2}{8\pi} ( \beta \mathcal{J} )^3 + \frac{13 - 112 Q_{0}^2 + 960 Q_{0}^4}{384\pi}( \beta \mathcal{J} )^5 +\dots\,, \quad \beta \mathcal{J} \ll 1\,.
 \end{align}
Substituting this expansion in Eq.~(\ref{largepmu}) we obtain 
\begin{align}
G_{\textrm{large}\, p}(\tau) =  G_{\textrm{HT}}(\tau) + O((\beta \mathcal{J})^4)\,,
\end{align}
where $G_{\textrm{HT}}(\tau)$ is the leading order of the high-temperature expansion for the Green's function:
 \begin{align}
G_{\textrm{HT}}(\tau) &= G_{0}(\tau) + J^{2} \int_{0}^{\beta} d\tau_{1}d\tau_{2} G_{0}(\tau-\tau_{1})G_{0}(\tau_{12})^{p/2}G_{0}(\beta-\tau_{12})^{p/2-1}G_{0}(\tau_{2}) +\dots  \notag\\
&=G_{0}(\tau) \left(1- \frac{1}{p}\left(\frac{\tau}{\beta}\Big(1-\frac{\tau}{\beta}\Big)+2Q_{0}\Big(\frac{\tau}{\beta}-Q_{0}-\frac{1}{2}\Big)\right)(\beta \mathcal{J})^{2}+\dots \right)\,. \label{GHmu}
\end{align}

\begin{figure}[h!]
    \centering
    \begin{subfigure}{0.48\textwidth}
        \centering
        \includegraphics[width=\textwidth]{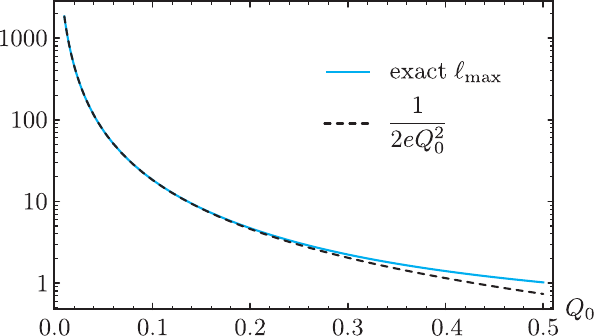}
        \caption{$\ell_{\textrm{max}}$ as function of $Q_{0}$}
        \label{fig:fig1}
    \end{subfigure}
    \quad
    \begin{subfigure}{0.48\textwidth}
        \centering
        \includegraphics[width=\textwidth]{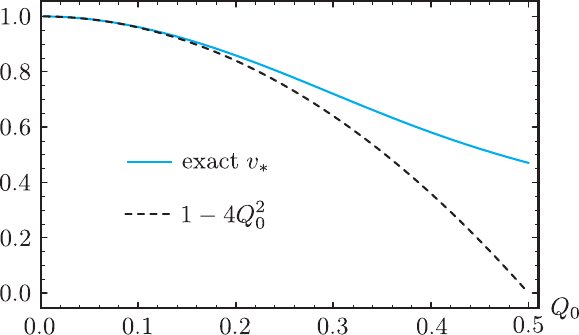}
        \caption{$v_{*}$ as function of $Q_{0}$}
        \label{fig:fig2}
    \end{subfigure}
    \caption{Blue lines represent exact values of $\ell_{\textrm{max}}$ and $v_{*}$  as functions of $Q_{0}$,  obtained by numerically finding the maximum of the function $\ell(v)$  defined in Eq.~(\ref{hdef}). The  black dashed lines correspond to the theoretical asymptotics.}
    \label{hmaxandv}
\end{figure}

For sufficiently large $\beta \mathcal{J}$ at fixed non-zero $\mu$, the transcendental equation Eq.~(\ref{betaJ_solution}) has no solution\footnote{We notice that this may be related to the complex SYK model undergoing a first-order phase transition at some critical $\mu$ at fixed $\beta\mathcal{J}$ \cite{PhysRevLett.120.061602, Complex_SYK_Phase_transition,Complex_SYK_Phase_transition2, Tikhanovskaya_2021} (see Appendix \ref{AppPhTr} for example).}. This contrasts with the $\mu=0$ case, where a solution exists for all $\beta\mathcal{J}$.
For the equation with nonzero $\mu$ to have a solution in the large $\beta\mathcal{J}$ regime, one must tune $Q_{0}$, and hence $\mu$ down as $\beta\mathcal{J}$ grows. To obtain an upper bound on $Q_{0}^{2}$ below which a solution to Eq.~\eqref{betaJ_solution} exists at large $\beta\mathcal{J}$, we take the limit $v\to1$ and expand the left-hand side of Eq.~\eqref{betaJ_solution} in this regime:
\begin{align}
\ell(v) \equiv \frac{\pi v}{\cos \frac{\pi v}{2}}e^{- 2\pi v \tan( \frac{\pi v}{2}) Q_{0}^{2} } \approx  \frac{2}{1-v} e^{-\frac{4Q_{0}^2}{1-v}} +\dots \,, \quad v \to 1 \,. \label{hdef}
\end{align}
We readily find that $\ell'(v)=0$ is solved by $v_{*} = 1-4Q_{0}^2$ and the maximum of $\ell(v)$ is $\ell_{\textrm{max}}\equiv \ell(v_{*})= 1/(2e Q_{0}^2)$\,. Therefore Eq.~(\ref{betaJ_solution})
has a solution if we satisfy
\begin{align}
Q_{0}^{2} \leq \frac{1}{2e \beta \mathcal{J}}, 
\end{align}
that we cited in Eq.~(\ref{mu_range})
\footnote{There are, in fact, two solutions. We select the one with $v<v_{*}$ since it is continuously connected to the solution at $\mu=0$.}.
In Figure \ref{hmaxandv} we plot the exact values of  $\ell_{\textrm{max}}$ and $v_{*}$ as functions of $Q_{0}$, obtained by numerically finding the maximum of the function $\ell(v)$ defined in Eq.~(\ref{hdef}), as well as the theoretical asymptotics derived above. We observe that for small $Q_{0}$ the asympotics rapidly approach the exact values.

For large $\beta \mathcal{J}$ and for $Q_{0}^{2} \leq 1/(2e \beta \mathcal{J})$ we can obtain a $1/\beta \mathcal{J}$ expansion for $v$, similar to the Majorana SYK model case \cite{Large_q}. Define the coefficient $\kappa_{0}$ by
 \begin{align}
Q_{0}^{2} = \frac{\kappa_{0}}{\beta \mathcal{J}}, \quad \kappa_{0} \leq \frac{1}{2e} \approx  0.184\,. \label{kappadef}
\end{align}
Then the first three terms of the expansion from Eq.~(\ref{betaJ_solution}) are 
\begin{align}
v = 1- \frac{v_{1}}{\beta \mathcal{J}} +\frac{v_{1}^2}{(\beta \mathcal{J})^{2}} - \frac{v_{1}^3 ((24 +\pi^2) v_{1}- 8\kappa_{0}(12-\pi^2)) }{24 (v_{1}-4\kappa_{0})(\beta \mathcal{J})^{3}} +\dots, \quad \beta \mathcal{J} \gg 1\,, \label{largeJmuexp}
\end{align}
where the coefficient $v_{1}$ is determined by the transcendental equation 
\begin{align}
v_{1} = 2 e^{-4\kappa_{0}/v_{1}}\,.
\end{align}
A solution for $v_{1}$ can be written via a chain sequence and to first order this gives $v_{1} \approx 2e^{-2\kappa_{0}}$. Using the expansion in Eq.~(\ref{largeJmuexp}), we can write $G_{\textrm{large}\,p}(\tau)$ to  leading order in $1/\beta \mathcal{J}$
\begin{align}
G_{\textrm{large}\,p}(\tau)= G_{0}(\tau) \frac{e^{-\frac{4\pi v}{p}\tan (\frac{\pi v}{2})Q_{0}   (\frac{\tau}{\beta}-\frac{1}{2})}}{(\frac{\beta \mathcal{J}}{\pi} \sin \frac{\pi \tau}{\beta} )^{2/p}} \left(1 -\frac{v_{1}}{p\beta \mathcal{J}}\Big(2 + \frac{\pi(1-2\tau/\beta) }{\tan \frac{\pi \tau}{\beta}}\Big)+\dots\right) \,, \label{lpG2}
\end{align}
where we have omitted  terms of order $1/(\beta \mathcal{J})^2$ and higher. We will see below that this result matches with the conformal solution. 

In the complex SYK model,  the conformal solution for the Green's function in the infrared regime acquires an additional asymmetry parameter $\mathscr{E}$ and takes the form on the interval $0<\tau <\beta$ \cite{Complex_SYK, Complex_SYK_Density_of_states}
\begin{align}
G_{c}(\tau) =  \frac{b^{1/p}}{(\frac{\beta J}{\pi} \sin \frac{\pi \tau}{\beta})^{2/p}}e^{2\pi \mathscr{E}(\frac{\tau}{\beta}-\frac{1}{2})}\,, \quad b = \frac{(1-\frac{2}{p})\sin(\frac{2\pi}{p})}{4\pi \cos(\frac{\pi}{p}+i\pi \mathscr{E}) \cos(\frac{\pi}{p}-i\pi \mathscr{E})}\,.
\end{align}
The asymmetry parameter $\mathscr{E}$ depends on $\mu$  and $\mathscr{E} = 0$ at $\mu=0$.
Replacing $J$ with $\mathcal{J}$ using Eq.~(\ref{betaJ_solution}), we can rewrite $G_{c}(\tau)$ in the form
\begin{align}
G_{c}(\tau) =  G_{0}(\tau) \Big(\frac{2p b}{1-4Q_{0}^2}\Big)^{1/p}\,
\frac{e^{(2\pi \mathscr{E}-\beta \mu )(\frac{\tau}{\beta}-\frac{1}{2})}}{(\frac{\beta \mathcal{J}}{\pi} \sin \frac{\pi \tau}{\beta})^{2/p}}\,. \label{Gcv2}
\end{align}
This expression agrees with the large $p$ result Eq.~(\ref{lpG2}) to leading order in large $\beta \mathcal{J}$, provided that
\begin{align}
2\pi \mathscr{E} = \beta \mu - \frac{1}{p}4\pi v \tan\Big(\frac{\pi v} {2}\Big) Q_{0} + O(1/p^2)\,, \label{Elp1}
\end{align}
and  this identification also implies    
\begin{align}
\Big(\frac{2p b}{1-4Q_{0}^2}\Big)^{1/p} = 1+ O(1/p^2)\,, 
\end{align}
therefore, there is full agreement between $G_{c}(\tau)$ and $G_{\textrm{large}\,p}(\tau)$ at leading order in both large $p$ and large $\beta \mathcal{J}$. 
This is true for the Majorana SYK model \cite{Large_q} at strong coupling $\beta{\mathcal J}\gg 1$, where correction to the conformal propagator also comes at the order $1/(p\beta{\mathcal J})$.
We show in Appendix~\ref{GpLuttRel} that, in the large $p$ limit, the asymmetry parameter in Eq.~(\ref{Elp1}) and the charge density $Q$ obey the Luttinger relation 
\cite{Complex_SYK_Density_of_states,Tikhanovskaya_2021}, providing an additional consistency check.

\subsection{Complex SYK grand potential in the large \texorpdfstring{$p$}{p} limit}

The main new result in this section is Eq.~(\ref{Omega}) for the thermodynamic grand potential $\Omega(\mu, T)$.
We start in the grand canonical ensemble. In order to calculate the grand potential 
$\Omega(\mu,T)=-T\log Z$ 
we use the $G\Sigma$ effective action  Eqs.~(\ref{partition_function},\ref{effective_action}) 
\begin{align}
\frac{\log Z}{N} =& \ln{\rm det}\left [\delta(\tau-\tau^{\prime})(\partial_{\tau}-\mu)-\Sigma(\tau,\tau^{\prime})\right]&& \nonumber\\
&-\beta \int_0^{\beta}d\tau \left[ \Sigma(\tau)G(\beta-\tau)+\Sigma(\beta-\tau)G(\tau)-\frac{J^2}{p} [G(\tau)]^{\frac{p}{2}}[G(\beta-\tau)]^{\frac{p}{2}} \right] \,.&&
\label{logZ}
\end{align} 
Following \cite{Large_q} we differentiate $\log Z/N$  with respect to coupling $J$
\begin{align}
 \frac{J\partial_{J}\log Z}{N}=&\frac{2J^2\beta}{p}\int d\tau [G(\tau)]^{\frac{p}{2}}[G(\beta-\tau)]^{\frac{p}{2}}=\frac{4\mathcal{J}^2\beta}{p^2\left(2\cosh\frac{\beta \mu}{2}\right)^2}\int d\tau e^{g_s(\tau)}\,,&&
\label{}
\end{align} 
where we used Eqs.~(\ref{G0G0}, \ref{betaJ_solution}). Using  
equation of motion (\ref{eqs_of_motion}) for $g_s$ we have
\begin{align}
\frac{J\partial_{J}\log Z}{N}=-\frac{4\beta\partial_{\tau}g_s(0)}{p^2\left(2\cosh\frac{\beta\mu}{2}\right)^2} \,.
\label{}
\end{align} 
Using the large $p$ solution Eq.~(\ref{g_solution}) 
\begin{align}
\partial_{\tau} g_s(\tau)=-\frac{2\pi v}{\beta}\tan\left(\frac{\pi v}{2}-\frac{\pi v\tau}{\beta}\right)
\label{}
\end{align} 
and turning $J\partial_{J} \rightarrow \mathcal{J}\partial_{\mathcal{J}} $ we obtain
\begin{align}
\frac{\mathcal{J}\partial_{\mathcal{J}} \log Z}{N}=\frac{2\pi v(1-4Q_0^2)}{p^2}\tan\left(\frac{\pi v}{2}\right),
\label{Jderivative}
\end{align}
where we represented $1-4Q_0^2=1/\cosh^2\left(\frac{\beta\mu}{2}\right)$ with $Q_0=\frac{1}{2}\tanh(\frac{\beta\mu}{2})$. Next we change the variables from $\mathcal{J}$ to $v$ and rewrite the derivative \cite{Large_q}
\begin{align}
\frac{\partial_v \log Z}{N}=\frac{\partial_v \mathcal{J}}{\mathcal{J}} \frac{\mathcal{J}\partial_{\mathcal{J}} \log Z}{N}\,.
\label{}
\end{align}
Using the relation between $\mathcal{J}$ and $v$ from Eq.~(\ref{betaJ_solution}) to calculate $\partial_v \mathcal{J}/\mathcal{J}$, we have 
\begin{align}
\frac{\partial_v \log Z}{N} =\frac{2\pi(1-4Q_0^2)}{p^2}\tan\left(\frac{\pi v}{2}\right)\left[1+\frac{\pi v}{2}\tan\left(\frac{\pi v}{2}\right) (1-4Q_{0}^2)
-\frac{(\pi v)^2 Q_0^2}{\cos^2\left(\frac{\pi v}{2}\right)}  \right]\,.&&
\label{}
\end{align}
Integrating over $v$ we obtain
\begin{align}
\frac{\log Z}{N} =\frac{2\pi v(1-4Q_0^2)}{p^2}\left[\tan\left(\frac{\pi v}{2}\right)\left(1- \pi v\tan\left(\frac{\pi v}{2}\right)Q_0^2\right)-\frac{\pi v}{4}\right]
\label{}
\end{align}
that gives the following grand potential
\begin{align}
\Omega(\mu,T) & =  -TN\ln \Big(2\cosh\Big(\frac{\beta\mu}{2}\Big)\Big)&&\nonumber\\
& -\frac{2\pi v TN}{\cosh^2(\frac{\beta\mu}{2})p^2}\left[\tan\left(\frac{\pi v}{2}\right)\left(1- \frac{\pi v}{4}\tan\left(\frac{\pi v}{2}\right)\tanh^2\left(\frac{\beta \mu}{2}\right)\right)-\frac{\pi v}{4}\right] \,,&&
\label{Omega}
\end{align}
where we fixed the integration constant using that $v=0$ at $\mathcal{J}=0$  and we should recover the free
value for the grand potential. Following \cite{Large_q} the energy $E=-\partial_{\beta}\log Z$ is found from the $\mathcal{J}$ derivative Eq.~(\ref{Jderivative})
\begin{align}
\frac{\mathcal{J}\partial_{\mathcal{J}} \log Z}{N}= -\beta E/N
\label{Energy}
\end{align}
and since the partition function depends on the combination $\beta\mathcal{J}$ and $\mathcal{J}\partial_{\mathcal{J}} =\beta\partial_{\beta}$. We obtain
\begin{align}
E=-\frac{2\pi v TN}{\cosh^2(\frac{\beta\mu}{2})p^2}\tan\left(\frac{\pi v}{2}\right)\,.
\label{FreeEnergy}
\end{align}

\section{Complex SYK model in the double scaling limit}
\label{section_DSSYK}

In this section we use double-scaled complex SYK model where $p\sim\sqrt{N}$ when $N\to\infty$ and $p\to \infty$, so that parameter 
\begin{align}
\lambda= \frac{p^2}{N} 
\end{align}
is kept fixed, and we take a limit of $\lambda\to 0$  
corresponding to classical expansion in the double-scaled SYK.
The Hamiltonian for the double-scaled complex SYK is given by Eq.~(\ref{complex_Hamiltonian}) with the variance \cite{Double_scaled_complex_SYK} 
\begin{align}
\left\langle J^{i_1\cdots i_{p/2}}_{j_1\cdots j_{p/2}}
J_{i_1\cdots i_{p/2}}^{j_1\cdots j_{p/2}} \right\rangle_J
=J^2\left(\begin{array}{c}
 N \\ p/2
\end{array}\right)^{-2}
\Big(2\cosh\frac{\beta\mu}{2}\Big)^{p} \,,
\end{align}
where we absorb the different normalization of fermions here \cite{Complex_SYK_Density_of_states} and in \cite{Double_scaled_complex_SYK} into the factor 
$(2\cosh\frac{\beta\mu}{2})^{p}$. Transforming from $J\rightarrow \bold{J}$ \cite{Double_scaled_complex_SYK}
\begin{align}
J^2=\frac{\bold{J}^2}{\lambda\cosh^2(\frac{\beta\mu}{2})}
\end{align}
and using the Stirling's formula we have an asymptotic behavior
\begin{align}
\left\langle J^{i_1\cdots i_{p/2}}_{j_1\cdots j_{p/2}}
J_{i_1\cdots i_{p/2}}^{j_1\cdots j_{p/2}} \right\rangle_J
=\bold{J}^2\frac{\pi \, p^{p-1}e^{-p}}{N^{p-1}}
\Big(\cosh\frac{\beta\mu}{2}\Big)^{p-2}\,,
\end{align}
which is the same as the variance in Eq.~(\ref{variance2}), since 
$\bold{J}\to \mathcal{J}$ with $\mathcal{J}$ given by eq.(\ref{betaJ_solution}) in the limit $\lambda\to 0$.
For a finite $\lambda$ the partition and two-point correlation functions have been calculated for complex double-scaled SYK in \cite{Double_scaled_complex_SYK}, 
where correlators are represented by a sum of cord diagrams which is solved by introducing a transfer matrix with integrals over energies. In the small $\lambda$ limit the integration problem is reduced to a classical action, 
in that the integrands become sharply peaked and dominated by a saddle point. We solve saddle point equations in the leading order of small $\lambda$ and obtain the grand potential and Green's function fluctuation.

In what follows we work with the chemical potential in the range
\begin{align}
(\beta\mu)^2 \leq \frac{2}{\beta\mathcal{J}},
\label{}
\end{align} 
or written through the charge density of free fermions $Q_0$ it is
\begin{align}
Q_0^2 \leq \frac{1}{8\beta\mathcal{J}}\,,
\label{rangeQ0}
\end{align} 
which is close to an estimate Eq.~(\ref{mu_range}) derived in the Section \ref{section_largepSYK}.
In this section we use convention for the chemical potential as in \cite{Double_scaled_complex_SYK} which is related to the chemical potential of the previous Section \ref{section_largepSYK} as $\mu_{\rm here}=-\beta\mu_{\rm there}/2$.

\subsection{Partition function and the grand potential}
The partition function of the complex SYK is the averaged thermodynamic partition function at the fixed chemical potential $\langle {\rm tr}\,{\rm e}^{-\beta H -2\mu Q} \rangle_{J}$.
The ensemble averaged grand canonical partition function 
was given in Ref.~\cite{Double_scaled_complex_SYK} in the form of resummed the moments of the Hamiltonian. In this section, we shall take the $\lambda \rightarrow 0$ limit of their result, and show that it agrees with our result of the large $p$ limit in Eq.~(\ref{Omega}).

The grand partition function in the double-scaled limit is \cite{Double_scaled_complex_SYK}
\begin{align}
Z(\beta,\mu) = \sum_{k=0}^{\infty} \frac{(-\beta)^k}{k!}m_k(\mu), \quad 
m_k(\mu)=\langle {\rm tr} \,H^{k}{\rm e}^{-2\mu Q}  \rangle_{J} ,
\label{}
\end{align}
The moments of the Hamiltonian operator, $m_k (\mu)$, for the double scaled complex SYK have been calculated using chord diagram technique \cite{Double_scaled_complex_SYK}
\begin{align}
m_k(\mu)=J^k \cosh(\mu)^{N-k p/2}\exp\left\{\frac{p^2}{8N}\left[k^2\sinh(\mu)^2-k(\sinh(2\mu)-1)\right]\right\}&&\nonumber\\
\times\sum_{CD}\exp\left\{-k_H\frac{p^2}{N}\cosh(\mu)^2\right\},&&
\label{momentH}
\end{align} 
where summation goes over the unoriented chord diagrams, $k_H$ is the number of intersections of H-chords, and $p/2$ is the number of fermions (and the same number of antifermions) comprising the Hamiltonian  -- the size of the Hamiltonian operator is $p$.

We take double scaled limit for the size of Hamilton operator
\begin{align}
p\sim \sqrt{N},\quad  {\rm as}\,\,  N\to \infty.
\label{double-scale1}
\end{align} 
Sum over the chord diagrams can be calculated as a matrix element of some transfer matrix, resulting in an integral over an angular variable $\theta$ that parametrizes energies. Therefore moments are represented through the density of states by an integral over the energies
\begin{align}
m_k(\mu)=\int dE \rho(E)E^k,
\label{standard_moment}
\end{align}
where relation between the angular variable $\theta$ and the energies is 
 \begin{align} 
E(\theta)=\frac{2J\gamma(\mu)\cos\theta}{\sqrt{1-q(\mu)}},
\label{energy-angle}
\end{align}
where $\gamma(\mu)$ is defined later and 
we introduced parameter $\lambda$ and $q$ associated with the size of Hamiltonian operator $p$
\begin{align}
q(\mu)={\rm e}^{-\lambda \cosh^2(\mu)}, \quad  \lambda=\frac{p^2}{N},
\label{}
\end{align}
From Eq.~(\ref{momentH}) moments of the Hamiltonian for the complex SYK in the double scaled limit have been calculated as an integral over angular variable $\theta$ \cite{Double_scaled_complex_SYK} 
 \begin{align}
m_k(\mu)=&\cosh(\mu)^{N-k p/2}\exp\left[\frac{p^2k^2\sinh^2(\mu)}{8N}\right]&&\nonumber\\
&\times\int_0^\pi
\frac{d\theta}{2\pi}(q(\mu),{\rm e}^{\pm 2i\theta};q(\mu))_{\infty}\left(\frac{2J\cos\theta}{\sqrt{1-q(\mu)}}{\rm e}^A \right)^k,&&
\label{}
\end{align} 
where the measure containing the $q$-Pochhammer symbols is the density of states. Here 
\begin{align}
A=-\frac{p^2}{8N}(\sinh(2\mu)-1)=-\frac{\lambda}{8}(\sinh(2\mu)-1),
\label{}
\end{align} 
and the $q$-Pochhammer symbol is defined by
\begin{align}
(x;q)_{\infty}=\prod_{k=0}^{\infty}(1-xq^k).
\label{}
\end{align} 
We used the shorthand notation for a product of three Pochhammer symbols
 $(q(\mu),{\rm e}^{\pm 2i\theta};q(\mu))_{\infty}=(q(\mu);q(\mu))_{\infty}\cdot ({\rm e}^{2i\theta};q(\mu))_{\infty}\cdot({\rm e}^{-2i\theta};q(\mu))_{\infty}$.
We do similar approximations and follow calculation scheme of \cite{Double_scaled_SYK_small_lambda}.
We consider the regime where $\lambda\to 0$, and we keep the energies finite.
It is convenient to rescale the coupling \cite{Double_scaled_SYK_small_lambda}
\begin{align}
J= \frac{\bold{J}}{\sqrt{\lambda\cosh^2(\mu)}}.
\label{coupling_rescale}
\end{align} 
In the limit $\lambda\to 0$ the partition function is
\begin{align}
Z(\beta,\mu)& =\cosh(\mu)^{N} \int_0^\pi
\frac{d\theta}{2\pi}(q(\mu),{\rm e}^{\pm 2i\theta};q(\mu))_{\infty}&& \nonumber\\
& ~~~~~~~~~~~\times\sum_{k=0}^{\infty}\frac{1}{k!}
\left(-\frac{2\beta\bold{J}\gamma(\mu)\cos\theta}{\lambda\cosh^2(\mu)}\right)^k\exp\left[\frac{\lambda}{8}\sinh^2(\mu)k^2\right],
\label{Zfunction}
\end{align} 
where
\begin{align}
\gamma(\mu)={\rm e}^{A-p\ln (\cosh(\mu))/2}\,,
\label{gamma0}
\end{align} 
with $p=\sqrt{\lambda N}$. 
In  this limit the energy becomes
 \begin{align} 
E(\theta)=\frac{2\bold{J}\gamma(\mu)\cos\theta}{\lambda\cosh^2(\mu)}.
\label{energy-angle2}
\end{align}
Due to  the term $\exp[\frac{\lambda}{8}\sinh^2(\mu)k^2]$ the moment does not have a standard form Eq.~(\ref{standard_moment}) as written through the energy density and the sum in the partition function diverges. 
However, we show that expression for $Z$ makes sense in the limit of small $\lambda$. To be consistent in the limit $\lambda\to 0$ we expand the exponent with $k^2$. Introducing 
\begin{align}
z=-\frac{2\beta\bold{J}\gamma(\mu)\cos\theta}{\lambda\cosh^2(\mu)},
\label{z}
\end{align} 
the sum in the partition function Eq.~(\ref{Zfunction}) is 
 \begin{align} 
\sum_{k=0}^{\infty}\frac{z^k}{k!}\left(1+\frac{\lambda}{8}\sinh^2(\mu)k^2+\frac{1}{2}\left(\frac{\lambda}{8}\right)^2\sinh^4(\mu)k^4\right)&&\nonumber\\
\approx
 {\rm e}^z\left(1+\frac{\lambda}{8}\sinh^2(\mu)z^2+\frac{1}{2}\left(\frac{\lambda}{8}\right)^2\sinh^4(\mu)z^4\right)&&\nonumber\\
 \approx \exp\left[z+\frac{\lambda}{8}\sinh^2(\mu)z^2\right],&&
\label{}
\end{align}
where we kept the leading order terms $\sim 1/\lambda$ in the exponent
\begin{align}
z+\frac{\lambda}{8}\sinh^2(\mu)z^2=-\frac{2\beta\bold{J}\gamma(\mu)\cos\theta}{\lambda\cosh^2(\mu)}+\frac{\tanh^2(\mu)\beta^2\bold{J}^2\gamma(\mu)^2\cos^2\theta}{2\lambda\cosh^2(\mu)}.
\label{gamma2}
\end{align} 
Expansion in Eq.~(\ref{gamma2}) is valid when
\begin{align}
z\geq \lambda\sinh^2(\mu)z^2,
\label{}
\end{align}
using Eq.~(\ref{z}) that gives the following range for the chemical potential
\begin{align}
\tanh^2(\mu)\leq \frac{1}{2\beta \bold{J}\gamma(\mu)},
\label{}
\end{align}
where we show later that $\mathcal{J}=\bold{J}\gamma(\mu)$. Rewriting it through $Q_0=\frac{1}{2}\tanh(\mu)$ we have
\begin{align}
Q_0^2\leq \frac{1}{8\beta \mathcal{J}},
\label{}
\end{align}
that is condition given in eq.(\ref{rangeQ0}).

At small $\lambda$ the Pochhammer symbol is approximated in the leading order
\begin{align}
(x;q(\mu))_{\infty}\approx  \exp \left[ -\frac{{\rm Li}_2(x)}{\lambda \cosh^2(\mu)}\right],
\label{Pocchammer_approx}
\end{align} 
where ${\rm Li}_2(x)$ is the dilogarithm. The limit $\lambda\to 0$ corresponds to the classical limit where we can use the saddle point approximation. 
Combining all the terms the partition function is
\begin{align}
Z(\beta,\mu)\approx \cosh(\mu)^{N} (q(\mu);q(\mu))_{\infty}  \int_0^\pi
\frac{d\theta}{2\pi} 
\exp\left[-\frac{f}{\lambda\cosh^2(\mu)}\right],
\label{Z0}
\end{align} 
where
\begin{align}
f={\rm Li}_2({\rm e}^{\pm 2i\theta})+2\beta\bold{J}\gamma(\mu)\cos\theta-\frac{1}{2}\beta^2\bold{J}^2\gamma(\mu)^2\tanh^2(\mu)\cos^2\theta,
\label{fZ}
\end{align} 
and $\pm$ means that the sum over the terms with plus and minus signs is taken. 
Using definition of the dilogarithm ${\rm Li}_2(z)=\sum_{k=1}^{\infty}\frac{z^k}{k^2}$, the derivative $\frac{d}{dx}{\rm Li}_2(x)=-\frac{\log(1-x)}{x}$, therefore 
$\frac{d}{dx}{\rm Li}_2({\rm e}^{\pm ix})=\mp i\log(1-{\rm e}^{\pm ix})$ and the saddle point equation is
\begin{align}
2\theta-\pi-\beta\bold{J}\gamma(\mu)\sin\theta\left[1-\frac{1}{2}\beta\bold{J}\gamma(\mu)\tanh^2(\mu)\cos\theta\right]=0.
\label{saddle_point}
\end{align} 
Defining as in \cite{Double_scaled_SYK_small_lambda}
\begin{align}
\theta=\frac{\pi}{2}+\frac{\pi v}{2},
\label{theta2}
\end{align} 
we solve the saddle point Eq.~(\ref{saddle_point})  in the limit of small $\lambda$ perturbatively. In the zeroth order it is
\begin{align}
\pi v-\beta\bold{J}\gamma(\mu)\cos\left(\frac{\pi v}{2}\right)=0,
\label{}
\end{align} 
that reproduces equation for $v$ in large $p$ calculations in \cite{Large_q} 
\begin{align}
\beta\bold{J}\,\gamma(\mu)=\frac{\pi v}{\cos\left(\frac{\pi v}{2}\right)},
\label{eqforv}
\end{align} 
with one difference that there is an additional factor $\gamma(\mu)$ defined in Eq.~(\ref{gamma0}).
In the first order the saddle point equation is
\begin{align}
\pi v-\beta\bold{J}\gamma(\mu)\cos\left(\frac{\pi v}{2}\right)\left[1+\frac{1}{2}\beta\bold{J}\gamma(\mu)\tanh^2(\mu)\sin\left(\frac{\pi v}{2}\right)\right]=0.
\label{}
\end{align} 
Using the zeroth order Eq.~(\ref{eqforv}), it becomes
\begin{align}
\beta\bold{J}\gamma(\mu)=\frac{\pi v}{\cos\left(\frac{\pi v}{2}\right)}\left[1-\frac{\pi v}{2}\tan\left(\frac{\pi v}{2}\right)\tanh^2(\mu)\right],
\label{equation_betaJ1}
\end{align} 
which can be written in the leading order 
\begin{align}
\beta\bold{J}\gamma(\mu)=\frac{\pi v}{\cos\left(\frac{\pi v}{2}\right)}{\rm e}^{-\frac{\pi v}{2}\tan\left(\frac{\pi v}{2}\right)\tanh^2(\mu)}.
\label{equation_betaJ2}
\end{align} 
Next we show that when the following identification is done this equation in double scale limit at small $\lambda$ coincides with Eq.~(\ref{betaJ_solution}) obtained in large $p$ calculations.
Here we used the same convention for the chemical potential as in \cite{Double_scaled_complex_SYK}, which is related to the chemical potential used in \cite{Complex_SYK_Density_of_states,Complex_SYK-Large_q,Tikhanovskaya_2021}
and in our large $p$ calculations in Section \ref{section_largepSYK} by $\mu_{\rm here}= -\beta\,\mu_{\rm there}/2$ (see a note in \cite{Double_scaled_complex_SYK}) . 
Taking into account that $Q_0=\frac{1}{2}\tanh(\frac{\beta\,\mu_{\rm there}}{2})$ and a correspondence $Q_0= -\frac{1}{2}\tanh(\mu_{\rm here})$, Eq.~(\ref{equation_betaJ2}) becomes
\begin{align}
\beta\bold{J}\gamma(\mu)=\frac{\pi v}{\cos\left(\frac{\pi v}{2}\right)}{\rm e}^{-2\pi v \tan\left(\frac{\pi v}{2}\right)Q_0^2},
\label{equation_betaJ3}
\end{align} 
which is the same as Eq.~(\ref{betaJ_solution}) with identification $\mathcal{J}=\bold{J}\gamma(\mu)$. Next we calculate the partition function at the saddle point solution.
Using a formula
\begin{align}
2\sum_{k=1}^{\infty}\frac{\cos kx}{k^2}={\rm Li}_2({\rm e}^{ix})+{\rm Li}_2({\rm e}^{-ix})=\frac{\pi^2}{3}-\pi x+\frac{x^2}{2},
\label{}
\end{align} 
and the saddle point solution $\theta$ Eq.~(\ref{theta2}) with $v$ defined by Eq.~(\ref{equation_betaJ1}) we have
\begin{align}
{\rm Li}_2({\rm e}^{2i\theta})+{\rm Li}_2({\rm e}^{-2i\theta})=2\sum_{k=1}^{\infty}\frac{\cos(2\theta k)}{k^2}=\frac{\pi^2}{3}-2\pi\theta+2\theta^2 = -\frac{\pi^2}{6}
+\frac{\pi^2v^2}{2}.
\label{Li2}
\end{align} 
Using the saddle point solution, the partition function Eq.~(\ref{Z0}) is $Z\sim (q(\mu);q(\mu))_{\infty}\exp\left[-\frac{f_{\rm saddle}}{\lambda\cosh^2(\mu)}\right]$ with $f$ defined by Eq.~(\ref{fZ}) is given by
\begin{align}
f_{\rm saddle}=-\frac{\pi^2}{6}+\frac{\pi^2v^2}{2}-2\pi v\tan\left(\frac{\pi v}{2}\right)\left[1-\frac{\pi v}{4}\tan\left(\frac{\pi v}{2}\right)\tanh^2(\mu)\right].
\label{f_saddle}
\end{align} 
In the leading order of small $\lambda$ the Pochhammer symbol is 
$(q(\mu);q(\mu))_{\infty}\sim \frac{1}{\sqrt{\lambda\cosh^2(\mu)}}\exp\left[-\frac{\pi^2}{6\lambda\cosh^2(\mu)}\right]$, that cancels with the similar term in $f_{\rm saddle}$ Eq.~(\ref{f_saddle}).
Therefore, we have for the partition function 
\begin{align}
Z\sim \cosh(\mu)^N \exp\left(-\frac{\pi^2v^2}{2\lambda\cosh^2(\mu)}+\frac{2\pi v\tan\left(\frac{\pi v}{2}\right)\left[1-\frac{\pi v}{4}\tan\left(\frac{\pi v}{2}\right)\tanh^2(\mu)\right]}{\lambda\cosh^2(\mu)} \right),
\label{Z}
\end{align} 
that gives for the grand potential $\Omega=-T\ln Z$ to the leading order
\begin{align}
\Omega =&-TN\ln(2\cosh(\mu))&&\nonumber\\
& -\frac{2\pi vT N}{\cosh^2(\mu)p^2}\left[\tan\left(\frac{\pi v}{2}\right)\left(1-\frac{\pi v}{4}\tan\left(\frac{\pi v}{2}\right)\tanh^2(\mu)\right)-\frac{\pi v}{4}\right],&&
\label{free_energy}
\end{align} 
where we used that $\lambda=\frac{p^2}{N}$ and an additional normalization term $-TN\ln 2$. The equation (\ref{free_energy}) agrees with our result in the large $p$ limit Eq.~(\ref{Omega})
when identification $\mu_{\rm here}=-\beta\mu_{\rm there}/2$ is done.

\subsection{Two-point function}

This section will take the results of Ref.~\cite{Double_scaled_complex_SYK} for the two-point correlator in the double-scaled limit, and show that its $\lambda \rightarrow 0$ limit agrees with Eq.~(\ref{g_solution}) obtained in the large $p$ limit.

We consider the two-point function $\langle {\rm tr}{\rm e}^{-\beta H}M(\tau)M(0)\rangle_{J,J_M}$, where the notation $\langle ...\rangle_{J,J_M}$ stands for the average over the ensemble of
both the Hamiltonian and operator couplings.
Writing the two point function for $\tau>0$ through the momentum $m_{k_1,k_2}(\mu)$ 
\begin{align}
G(\tau)=-\frac{1}{Z(\beta,\mu)}\sum_{k_1,k_2=0}^{\infty}\frac{(-\tau)^{k_1}}{k_1!}\frac{(\tau-\beta)^{k_2}}{k_2!}m_{k_1,k_2}(\mu),
\label{2point}
\end{align} 
where the moments are defined as
\begin{align}
m_{k_1,k_2}(\mu)={\rm e}^{\frac{2\tau\mu(\bar{p}_M-p_M)}{\beta}}\langle {\rm tr}MH^{k_1}\bar{M}H^{k_2}\exp\left[-\frac{\mu}{2}\sum_{i=1}^N[\bar{\psi}_i,\psi^i]\right]\rangle_{J,J_M}\,.
\label{}
\end{align} 
Using the rules of evaluating chord diagrams in the large $N$ limit, the moments for two-point function of double scale complex SYK have been found \cite{Double_scaled_complex_SYK}
\begin{align}
& m_{k_1,k_2}(\mu)=J_M^2J^k{\rm e}^{\mu(1-2\frac{\tau}{\beta})(p_M-\bar{p}_M)}\cosh(\mu)^{N-kp/2-p_M-\bar{p}_M}\nonumber\\
&~~ \times\exp\left\{
\frac{p^2}{8N}\left[k^2\sinh(\mu)^2-k(\sinh(2\mu)-1)\right]
 -\frac{pp_M}{4N}\left[k_1(1-{\rm e}^{2\mu})+k_2(1-{\rm e}^{-2\mu})\right]\right.\nonumber\\
&~~~~~~~~~~~~\left.-\frac{p\bar{p}_M}{4N}\left[k_1(1-{\rm e}^{-2\mu})+k_2(1-{\rm e}^{2\mu})\right]
+\frac{p_M\bar{p}_M}{N}{\rm e}^{2\mu}
\right\}\nonumber\\
&~~~~~~~~~~~~~~~~~\times\sum_{CD}\exp\left\{-k_H\frac{p^2}{N}\cosh(\mu)^2-k_{HM}\frac{p(p_M+\bar{p}_M)}{N}\cosh(\mu)^2
\right\},
\label{}
\end{align} 
where summation goes over chord diagrams where $k_H$ is the number of intersections of H-chords and 
 $k_{HM}$ is the number of intersections of the M -chord with H-chords, again $p$ is the size of the Hamiltonian operator, and 
 $p_M$ is the number of fermions and $\bar{p}_M$ is the number of antifermions in an operator M. We refer to a sum
 $p_M+\bar{p}_M$ as the size of the operator M, and the difference $p_M-\bar{p}_M$ determines the charge.
We take the size of the operator M to be double scaled as well as double scaling for the size of Hamiltonian in Eq.~(\ref{double-scale1}) 
\begin{align}
p_M,\, \bar{p}_M\sim \sqrt{N},\quad {\rm as}\,\, N\to\infty
\label{double-scale2}
\end{align} 

In the double-scaled complex SYK the moment for the two-point function is given by an integral over two angle variables associated with two energies as in Eq.~(\ref{energy-angle}) \cite{Double_scaled_complex_SYK}
\begin{align}
 & m_{k_1,k_2}(\mu)  =J_M^2{\rm e}^{\mu(1-2\tau/\beta)(p_M-\bar{p}_M)}\cosh(\mu)^{N-kp/2-p_M-\bar{p}_M}\nonumber\\
 & ~~~\times\exp\left[\frac{p^2k^2\sinh^2(\mu)}{8N}+\frac{p_M\bar{p}_M}{N}{\rm e}^{2\mu}\right]
\int_0^{\pi}\prod_{j=1,2}\left[\frac{d\theta_j}{2\pi}(q(\mu),{\rm e}^{\pm 2i\theta_j};q(\mu))_{\infty}\right]\nonumber\\
& ~~~~\times\left(\frac{2J\cos\theta_1}{\sqrt{1-q(\mu)}}{\rm e}^{A+B_1}\right)^{k_1}\left(\frac{2J\cos\theta_2}{\sqrt{1-q(\mu)}}{\rm e}^{A+B_2}\right)^{k_2}
\frac{(\tilde{q}(\mu)^2;q(\mu))_{\infty}} {(\tilde{q}(\mu){\rm e}^{i(\pm \theta_1\pm\theta_2)};q(\mu))_{\infty}}\,,
\label{2point_momentum}
\end{align} 
where $k=k_1+k_2$, and again there is a product of several Pochhammer symbols. Here we introduced along with $q$ associated with the size of the Hamiltonian operator a corresponding parameter $\tilde{q}$ associated with the size (number of fermions) in operator $M$
\begin{align}
q(\mu)&={\rm e}^{-\lambda\cosh^2(\mu)}, \quad \lambda=\frac{p^2}{N},&&\nonumber\\
\tilde{q}(\mu)&={\rm e}^{-\tilde{\lambda}\cosh^2(\mu)}, \quad \tilde{\lambda}=\frac{p(p_M+\bar{p}_M)}{N}.&&
\label{}
\end{align} 
Also we introduced 
\begin{align}
A=-\frac{p^2}{8N}(\sinh(2\mu)-1), \quad
B_1= -\frac{pp_M}{4N}(1-{\rm e}^{2\mu})-\frac{p\bar{p}_M}{4N}(1-{\rm e}^{-2\mu}),&&\nonumber\\
B_2= -\frac{pp_M}{4N}(1-{\rm e}^{-2\mu})-\frac{p\bar{p}_M}{4N}(1-{\rm e}^{2\mu}).&&
\label{coefficients}
\end{align}
Let us use the sum 
$(p_M+\bar{p}_M)$ which is proportional to the size of operator $M$ and the difference $(p_M-\bar{p}_M)$ which is proportional to the total charge in the corresponding $\lambda$'s parameters 
\begin{align}
\tilde{\lambda}=\frac{p(p_M+\bar{p}_M)}{N},\quad
\lambda_{\mu}=\frac{p(p_M-\bar{p}_M)}{N},
\label{}
\end{align}
where $\lambda_{\mu}=0$ at zero charge.
Therefore we can rewrite Eq.~(\ref{coefficients})
\begin{align}
A &= -\frac{\lambda}{8}(\sinh(2\mu)-1), \quad
B= -\frac{\tilde{\lambda}}{4}(1-\cosh(2\mu)),
&&\nonumber\\
B_1&= B+\frac{\lambda_{\mu}}{4}\sinh(2\mu), \quad
B_2= B-\frac{\lambda_{\mu}}{4}\sinh(2\mu).&& 
\label{AB}
\end{align}
We do similar approximations and follow calculation scheme of \cite{Double_scaled_SYK_small_lambda}.
We consider all $\lambda$'s to be small, so that operators $M$ are light and there is no backreaction. 
Substituting the momentum Eq.~(\ref{2point_momentum}) in the two-point function Eq.~(\ref{2point}), $G=\tilde{G}/Z$,
we have in the limit $\lambda\to 0$
\begin{align}
\tilde{G}(\tau)  =&J_M^2{\rm e}^{\mu(1-2\tau/\beta)(p_M-\bar{p}_M)}\cosh(\mu)^{N-p_M-\bar{p}_M}\exp\left[\frac{p_M\bar{p}_M}{N}{\rm e}^{2\mu}\right]\nonumber\\
& \times\int_0^{\pi}\prod_{j=1,2}\left[\frac{d\theta_j}{2\pi}(q(\mu),{\rm e}^{\pm 2i\theta_j};q(\mu))_{\infty}\right] 
 \frac{(\tilde{q}(\mu)^2;q(\mu))_{\infty}} {(\tilde{q}(\mu){\rm e}^{i(\pm \theta_1\pm\theta_2)};q(\mu))_{\infty}}\nonumber\\
& \times\sum_{k_1k_2}\frac{1}{k_1!}
\left( -\frac{2\tau\bold{J}\tilde{\gamma}_1(\mu)\cos\theta_1}{\lambda\cosh^2(\mu)}\right)^{k_1}
\frac{1}{k_2!}
\left(-\frac{2(\beta-\tau)\bold{J}\tilde{\gamma}_2(\mu)\cos\theta_2}{\lambda\cosh^2(\mu)}\right)^{k_2}\nonumber\\
& ~~~~~~~~~~\times\exp\left[\frac{\lambda}{8}\sinh^2(\mu)(k_1+k_2)^2\right],
\label{Greens_function}
\end{align} 
where $k=k_1+k_2$ and we rescaled the coupling constant $J$ Eq.~(\ref{coupling_rescale}) and introduced
\begin{align}
\gamma_1(\mu)&={\rm e}^{A+B_1-p\ln(\cosh(\mu))/2}=\tilde{\gamma}(\mu)\,{\rm e}^{\frac{\lambda_{\mu}}{4}\sinh(2\mu)},&&\nonumber\\
\gamma_2(\mu)&={\rm e}^{A+B_2-p\ln(\cosh(\mu))/2}=\tilde{\gamma}(\mu)\,{\rm e}^{-\frac{\lambda_{\mu}}{4}\sinh(2\mu)},&&
\label{}
\end{align}  
with
\begin{align}
\tilde{\gamma}(\mu)={\rm e}^{A+B-p\ln(\cosh(\mu))/2}=\gamma(\mu){\rm e}^{B}=\gamma(\mu){\rm e}^{-\frac{\tilde{\lambda}}{4}(1-\cosh(2\mu))},
\label{tildegamma}
\end{align}  
where $\gamma(\mu)$ was defined for the partition function in Eq.~(\ref{gamma0}).
Due to  the term $\exp[\frac{\lambda}{8}\sinh^2(\mu)k^2]$ the Green's function diverge. However,  in the limit of small $\lambda$ we expand the exponent with $k^2$, the same way as we did for the partition function. We  introduce 
\begin{align}
z_1=-\frac{2\tau\bold{J}\gamma_1(\mu)\cos\theta_1}{\lambda\cosh^2(\mu)}
,\quad z_2=-\frac{2(\beta-\tau)\bold{J}\gamma_2(\mu)\cos\theta_2}{\lambda\cosh^2(\mu)}.
\label{gamma}
\end{align} 
Keeping the leading in $1/\lambda$ terms, the sum in the Green's function Eq.~(\ref{Greens_function}) is
\begin{align}
\sum_{k_1,k_2=0}^{\infty}\frac{z_1^{k_1}}{k_1!}\cdot \frac{z_2^{k_2}}{k_2!}\left(1+\frac{\lambda}{8}\sinh^2(\mu)(k_1+k_2)^2\right)\approx {\rm e}^{z_1+z_2}\left(1+\frac{\lambda}{8}\sinh^2(\mu)(z_1+z_2)^2\right)&&\nonumber\\
\approx \exp\left[z_1+z_2+\frac{\lambda}{8}\sinh^2(\mu)(z_1+z_2)^2\right], &&
\label{}
\end{align}  
with
\begin{align}
z_1+z_2+\frac{\lambda}{8}\sinh^2(\mu)(z_1+z_2)^2=-\frac{2\bold{J}\tau\gamma_1(\mu)\cos\theta_1+2\bold{J}(\beta-\tau)\gamma_2(\mu)\cos\theta_2}{\lambda\cosh^2(\mu)}&&\nonumber\\
\qquad +\frac{\tanh^2(\mu)\left(\bold{J}\tau\gamma_1(\mu)\cos\theta_1+\bold{J}(\beta-\tau)\gamma_2(\mu)\cos\theta_2\right)^2}{2\lambda\cosh^2(\mu)}.&&
\label{}
\end{align}  
Using the leading asymptotics for Pochhammer symbols Eq.~(\ref{Pocchammer_approx}) the two-point function has $1/\lambda$ factor in the exponent 
\begin{align}
\tilde{G}=&J_M^2{\rm e}^{\mu(1-2\tau/\beta)(p_M-\bar{p}_M)}\cosh(\mu)^{N-p_M-\bar{p}_M}\exp\left[\frac{p_M\bar{p}_M}{N}{\rm e}^{2\mu}\right]&&\nonumber\\
&\times(q(\mu);q(\mu))_{\infty}^2(\tilde{q}(\mu)^2;q(\mu))_{\infty}
\int_{0}^{\pi}\frac{d\theta_1d\theta_2}{(2\pi)^2}\exp\left[-\frac{f}{\lambda\cosh^2(\mu)}\right],&&
\label{G}
\end{align} 
corresponding to the classical approximation. Here the exponent is given by
\begin{align}
&f={\rm Li}_2({\rm e}^{\pm2i\theta_1})+{\rm Li}_2({\rm e}^{\pm2i\theta_2})
-{\rm Li}_2(\tilde{q}(\mu){\rm e}^{i(\pm \theta_1\pm\theta_2)})
+2\bold{J}\tau\gamma_1(\mu)\cos\theta_1&&\nonumber\\
&\;\;+2\bold{J}(\beta-\tau)\gamma_2(\mu)\cos\theta_2
-\frac{1}{2} \left(\bold{J}\tau\gamma_1(\mu)\cos\theta_1+\bold{J}(\beta-\tau)\gamma_2(\mu)\cos\theta_2\right)^2 \tanh^2(\mu)\,.
\label{f}
\end{align} 
This allows to use saddle point approximation to evaluate the integrals. The saddle point equations are
\begin{align}
2\theta_1-\pi +\tan^{-1}\left[\frac{\sin(\theta_1+\theta_2)}{{\rm e}^{\tilde{\lambda}\cosh^2(\mu)}-\cos(\theta_1+\theta_2)}\right]
+\tan^{-1}\left[\frac{\sin(\theta_1-\theta_2)}{{\rm e}^{\tilde{\lambda}\cosh^2(\mu)}-\cos(\theta_1-\theta_2)}\right]&&\nonumber\\
=\bold{J}\tau\gamma_1(\mu)\sin\theta_1\left(1-\frac{1}{2}\left(\bold{J}\tau\gamma_1(\mu)\cos\theta_1+\bold{J}(\beta-\tau)\gamma_2(\mu)\cos\theta_2\right)\tanh^2(\mu)\right)&& 
\label{saddle1}\\
2\theta_2-\pi +\tan^{-1}\left[\frac{\sin(\theta_1+\theta_2)}{{\rm e}^{\tilde{\lambda}\cosh^2(\mu)}-\cos(\theta_1+\theta_2)}\right]
-\tan^{-1}\left[\frac{\sin(\theta_1-\theta_2)}{{\rm e}^{\tilde{\lambda}\cosh^2(\mu)}-\cos(\theta_1-\theta_2)}\right]&&\nonumber\\
=\bold{J}(\beta-\tau)\gamma_2(\mu)\sin\theta_2\left(1-\frac{1}{2}\left(\bold{J}\tau\gamma_1(\mu)\cos\theta_1+\bold{J}(\beta-\tau)\gamma_2(\mu)\cos\theta_2\right)\tanh^2(\mu)\right)&&
\label{saddle2}
\end{align} 
where we used definition of the dilogarithm ${\rm Li}_2(z)=\sum_{k=1}^{\infty}\frac{z^k}{k^2}$, the derivative $\frac{d}{dx}{\rm Li}_2(x)=-\frac{\log(1-x)}{x}$, therefore 
$\frac{d}{dx}{\rm Li}_2({\rm e}^{ix})=-i\log(1-{\rm e}^{ix})$ and $\tan^{-1}(x)=\frac{i}{2}[\log(1-ix)-\log(1+ix)]$. 
Introducing the following ansatz as in \cite{Double_scaled_SYK_small_lambda} and working to the order $O(\tilde{\lambda})$
\begin{align}
\theta_1=\theta+\alpha\tilde{\lambda}\cosh^2(\mu), \quad \theta_2=\theta-\alpha\tilde{\lambda}\cosh^2(\mu),
\label{theta1,2}
\end{align}  
we get to the leading order from Eqs.~(\ref{saddle1},\ref{saddle2}) 
\begin{align}
2\theta-\pi-2\bold{J}\tau\tilde{\gamma}(\mu)\sin\theta\left(1-\frac{1}{2}\bold{J}\beta\tilde{\gamma}(\mu)\tanh^2(\mu)\cos\theta\right)+2\tan^{-1}(2\alpha)=0&&\label{saddlepointeqs1}\\
2\theta-\pi-2\bold{J}(\beta-\tau)\tilde{\gamma}(\mu)\sin\theta\left(1-\frac{1}{2}\bold{J}\beta\tilde{\gamma}(\mu)\tanh^2(\mu)\cos\theta\right)-2\tan^{-1}(2\alpha)=0&&
\label{saddlepointeqs2}
\end{align} 
where $\alpha$ enters to the order $O(\tilde{\lambda}^0)$ because of the singular behavior which arises due to the same saddle point value of $\theta_1$ and $\theta_2$ at leading order. 
Sum of two equations (\ref{saddlepointeqs1},\ref{saddlepointeqs2}) gives the same equation, except for the difference between $\gamma(\mu)$ and $\tilde{\gamma}(\mu)$, 
as in the case of the partition function Eq.~(\ref{saddle_point})
\begin{align}
2\theta-\pi-\beta\bold{J}\tilde{\gamma}(\mu)\sin\theta\left[1-\frac{1}{2}\bold{J}\beta\tilde{\gamma}(\mu)\tanh^2(\mu)\cos\theta\right]=0.
\label{sum}
\end{align}  
Using parametrization 
\begin{align}
\theta=\frac{\pi}{2}+\frac{\pi\tilde{v}}{2}\,,
 \label{theta}
\end{align}
and solving Eq.~(\ref{sum}) perturbatively, we get
\begin{align}
\beta\bold{J}\tilde{\gamma}(\mu)=\frac{\pi\tilde{v}}{\cos\frac{\pi\tilde{v}}{2}}\left[1-\frac{\pi\tilde{v}}{2}\tan\left(\frac{\pi\tilde{v}}{2}\right)\tanh^2(\mu)\right],
\label{betaJv0}
\end{align}
which to the leading order is similar to Eq.~(\ref{eqforv}) for the partition function
\begin{align}
\beta\bold{J}\tilde{\gamma}(\mu)=\frac{\pi\tilde{v}}{\cos\frac{\pi\tilde{v}}{2}}{\rm e}^{-\frac{\pi\tilde{v}}{2}\tan\left(\frac{\pi\tilde{v}}{2}\right)\tanh^2(\mu)}.
\label{betaJv}
\end{align}
We distinguish $\tilde{v}$ satisfying equation (\ref{betaJv}) with $\tilde{\gamma}(\mu)$.
Difference of two equations (\ref{saddlepointeqs1},\ref{saddlepointeqs2}) is
\begin{align}
\beta\bold{J}\tilde{\gamma}(\mu)\sin\theta\left[1-\frac{1}{2}\bold{J}\beta\tilde{\gamma}(\mu)\tanh^2(\mu)\cos\theta\right]\left(1-\frac{2\tau}{\beta}\right)+2\tan^{-1}(2\alpha)=0.
\label{}
\end{align}  
Using the relation between $\tilde{v}$ and $\beta\bold{J}\tilde{\gamma}(\mu)$ as defined in equation for the sum (\ref{betaJv0})  we get 
\begin{align}
\pi\tilde{v}\left[1-2\left(\frac{\pi\tilde{v}}{2}\right)^2\tan^2\left(\frac{\pi\tilde{v}}{2}\right)\tanh^4(\mu)\right] (1-\frac{2\tau}{\beta})=-2\tan^{-1}(2\alpha), 
\label{alpha0}
\end{align}  
that gives to the leading order
\begin{align}
\alpha=-\frac{1}{2}\tan\left[\frac{\pi \tilde{v}}{2}(1-\frac{2\tau}{\beta})\right],
\label{alpha}
\end{align} 
that coincides with the case of zero charge, $\mu=0$. In what follows we substitute the saddle point solutions to the Green's function.
We use the formulas
\begin{align}
\sum_{k=1}^{\infty} \frac{\cos(kx)}{k^2}=&\frac{1}{2}\left( {\rm Li}({\rm e}^{ix})+{\rm Li}({\rm e}^{-ix})\right)=\frac{\pi^2}{6}-\frac{\pi x}{2}+\frac{x^2}{4},&&\label{formula1}\\
\lim_{a\to 0}\sum_{k=1}^{\infty}{\rm e}^{-ak}\frac{\cos(kx)}{k^2}=& \frac{\pi^2}{6}-\frac{\pi x}{2}+\frac{x^2}{4}+a\log[2\sin\left(\frac{x}{2}\right)],&&\label{formula2}\\
\lim_{a\to 0}{\rm Li}_2({\rm e}^{-ax})=&\frac{\pi^2}{6}+xa\left(\log a+\log x -1\right)-\frac{x^2a^2}{4}.
\label{formula3}
\end{align}  
Using Eqs.~(\ref{formula1}-\ref{formula3}) and substituting the saddle point solution Eqs.~(\ref{theta1,2},\ref{theta}) with Eqs.~(\ref{betaJv0},\ref{alpha}) into the terms of the exponent $f$ Eq.~(\ref{f}) 
that enters the Green's function Eq.~(\ref{G}), we have to the leading order
\begin{align}
{\rm Li}_2({\rm e}^{\pm 2i\theta_1})+{\rm Li}_2({\rm e}^{\pm 2i\theta_2})
\approx& \frac{2\pi^2}{3}-4\pi\theta+4\theta^2=-\frac{\pi^2}{3}+\pi^2v^2,&&\label{first}\\
-{\rm Li}_2(\tilde{q}(\mu){\rm e}^{\pm i(\theta_1+\theta_2)})\approx& -\frac{\pi^2}{3}+2\pi\theta-2\theta^2 -2\tilde{\lambda}\cosh^2(\mu)\log[2\sin\theta]&&\nonumber\\
=&\frac{\pi^2}{6}-\frac{\pi^2 v^2}{2}
-2\tilde{\lambda}\cosh^2(\mu)\log\left[2\cos\left(\frac{\pi v}{2}\right)\right],&&\label{second}\\
-{\rm Li}_2(\tilde{q}(\mu){\rm e}^{\pm i(\theta_1-\theta_2)})\approx& -\frac{\pi^2}{3}+\tilde{\lambda}\cosh^2(\mu)\left[2-2\log (\tilde{\lambda}\cosh^2(\mu))-\log(1+4\alpha^2)\right]&&\nonumber\\
&+4\alpha\tilde{\lambda}\cosh^2(\mu)\tan^{-1}(2\alpha)=-\frac{\pi^2}{3}&&\nonumber\\
&+\tilde{\lambda}\cosh^2(\mu)\left[2-2\log(\tilde{\lambda}\cosh^2(\mu))+\log\Big(\cos^2(\frac{\pi v}{2}(1-\frac{2\tau}{\beta}))\Big)\right]&&\nonumber\\
&+2\tilde{\lambda}\cosh^2(\mu)\tan\left(\frac{\pi v}{2}(1-\frac{2\tau}{\beta})\right)\times\left(\frac{\pi v}{2}(1-\frac{2\tau}{\beta})\right),&&\label{third}
\end{align} 
where $\tilde{v}\approx v$ to this order,
and 
\begin{align}
&2\bold{J}\tau\gamma_1(\mu)\cos\theta_1+2\bold{J}(\beta-\tau)\gamma_2(\mu)\cos\theta_2-\frac{1}{2}\left(\bold{J}\tau\gamma_1(\mu)\cos\theta_1+\bold{J}(\beta-\tau)\gamma_2(\mu)\cos\theta_2\right)^2\tanh^2(\mu)&&\nonumber\\
&\quad \approx 2\bold{J}\beta\gamma(\mu){\rm e}^{B}\cos\theta
+2\alpha\tilde{\lambda}\cosh^2(\mu)\bold{J}\beta\gamma(\mu)\sin\theta\left(1-\frac{2\tau}{\beta}\right)&&\nonumber\\
&\quad\quad -\frac{\lambda_{\mu}}{2}\sinh(2\mu)\bold{J}\beta\gamma(\mu)\cos\theta\left(1-\frac{2\tau}{\beta}\right) -\frac{1}{2}\bold{J}^2\beta^2\gamma^2(\mu)\tanh^2(\mu)\cos^2\theta&&\nonumber\\
&\quad=-2\pi v\tan\left(\frac{\pi v}{2}\right)\left[1-\frac{\tilde{\lambda}}{4}(1-\cosh(2\mu))\right] +\frac{\pi^2 v^2}{2}\tan^2\left(\frac{\pi v}{2}\right)\tanh^2(\mu)&&\nonumber\\
&\quad\quad-2\tilde{\lambda}\cosh^2(\mu)\tan\left(\frac{\pi v}{2}(1-\frac{2\tau}{\beta})\right)\times\left(\frac{\pi v}{2}(1-\frac{2\tau}{\beta})\right)
&&\nonumber\\
&\quad\quad+\lambda_{\mu}\sinh(2\mu)\tan\left(\frac{\pi v}{2}\right)\times \left(\frac{\pi v}{2}(1-\frac{2\tau}{\beta})\right),
\label{fourth}
\end{align} 
where we used Eq.~(\ref{tildegamma}) for relation between $\tilde{\gamma}(\mu)$ and $\gamma(\mu)$ with $B$ given by Eq.~(\ref{AB}), and Eq.~(\ref{equation_betaJ1}) that satisfies $v$.
In Eq.~(\ref{fourth}) to the leading order we neglected terms $\sim \tilde{\lambda}\tanh^2(\mu)$ and $\sim \lambda_{\mu}\tanh^2(\mu)$.
Combining all the terms, adding Eqs.~(\ref{first}-\ref{fourth}), we get $f$ Eq.~(\ref{f}) evaluated at the saddle point  
\begin{align}
f_{\textrm{saddle}}=&-\frac{\pi^2}{2}+\frac{\pi^2 v^2}{2}\left(1+\tan^2\left(\frac{\pi v}{2}\right)\tanh^2(\mu)\right)&&\nonumber\\
&-2\pi v\tan\left(\frac{\pi v}{2}\right)\left[1-\frac{\tilde{\lambda}}{4}(1-\cosh(2\mu))\right] &&\nonumber\\
&+\tilde{\lambda}\cosh^2(\mu)\left[2-2\log(2\tilde{\lambda}\cosh^2(\mu))+\log\left(\frac{\cos^2\left(\frac{\pi v}{2}(1-\frac{2\tau}{\beta})\right)}{\cos^2\left(\frac{\pi v}{2}\right)}\right)\right]&&\nonumber\\
&+\lambda_{\mu}\sinh(2\mu)\tan\left(\frac{\pi v}{2}\right)\times \left(\frac{\pi v}{2}(1-\frac{2\tau}{\beta})\right).
\label{fsaddle2}
\end{align} 
Note that the linear in time term, $\sim\tilde{\lambda}\cosh^2(\mu)(1-\frac{2\tau}{\beta})$, cancels, but the linear in time term $\sim\lambda_{\mu}\sinh(2\mu)(1-\frac{2\tau}{\beta})$ arises due to nonzero charge. 
To the leading order  the Pochhammer symbols in the Green's function $\tilde{G}$ Eq.~(\ref{G}) are
\begin{align}
(q(\mu);q(\mu))^2_{\infty}&\sim \frac{1}{\lambda\cosh^2(\mu)}\exp\left[-\frac{\pi^2}{3\lambda\cosh^2(\mu)}\right],&&\\
(\tilde{q}^2(\mu);q(\mu))_{\infty}&\sim \exp\left[-\frac{\pi^2}{6\lambda\cosh^2(\mu)}+\frac{\tilde{\lambda}}{\lambda}\left(2-2\log(2\tilde{\lambda}\cosh^2(\mu))\right)\right]. 
\label{}
\end{align} 
The partition function Eq.~(\ref{Z})
is 
\begin{align}
Z\sim \exp\left[-\frac{1}{\lambda\cosh^2(\mu)}\left(\frac{\pi^2 v^2}{2}+\frac{\pi^2 v^2}{2}\tan^2\left(\frac{\pi v}{2}\right)\tanh^2(\mu)-2\pi v \tan\left(\frac{\pi v}{2}\right)\right)\right].
\label{}
\end{align} 
Therefore the normalized two-point function, $G=\tilde{G}/Z$ where
$\tilde{G}\sim \exp[-\frac{f_{\textrm{saddle}}}{\lambda\cosh^2(\mu)}]$ with $f_{\textrm{saddle}}$ given by Eq.~(\ref{fsaddle2}), becomes
\begin{align}
G = \left(\frac{\cos^2\left(\frac{\pi v}{2}\right)}{\cos^2\left(\frac{\pi v}{2}(1-\frac{2\tau}{\beta})\right)}\right)^{\frac{\tilde{\lambda}}{\lambda}} \times
\left(\exp\left[-\pi v\tan\left(\frac{\pi v}{2}\right)\tanh(\mu)
\times(1-\frac{2\tau}{\beta})\right]\right)^{\frac{\lambda_{\mu}}{\lambda}}&&\nonumber\\
\times \left(\exp\left[\pi v\tan\left(\frac{\pi v}{2}\right)\tanh^2(\mu)\right]\right)^{\frac{\tilde{\lambda}}{\lambda}}.&&
\label{GG}
\end{align} 
Making connection with \cite{Large_q}, we can write $G={\rm e}^{g(\tau)}$. All the lambda's are small, $\frac{\tilde{\lambda}}{\lambda}\to 1$ and $\frac{\tilde{\lambda}_{\mu}}{\lambda}\to 1$,  we have from Eq.~(\ref{GG})
\begin{align}
g(\tau) =\frac{2\pi v}{\beta}\tan\left(\frac{\pi v}{2}\right)\tanh(\mu)\left(\tau-\frac{\beta}{2}(1-\tanh(\mu))\right) +\log\left(\frac{\cos^2\left(\frac{\pi v}{2}\right)}{\cos^2\left(\frac{\pi v}{2}(1-\frac{2\tau}{\beta})\right)}\right), 
\label{g_solution0}
\end{align} 
where there is the following relation between parameter $v$ and $\beta\bold{J}$ Eq.~(\ref{equation_betaJ2})
\begin{align}
\beta\bold{J}\gamma(\mu)=\frac{\pi v}{\cos\left(\frac{\pi v}{2}\right)}{\rm e}^{-\frac{\pi v}{2}\tan\left(\frac{\pi v}{2}\right)\tanh^2(\mu)}\,.
\label{betaJ_solution0}
\end{align} 
As noted in \cite{Double_scaled_complex_SYK} at page 5, we used the convention for the chemical potential of \cite{Double_scaled_complex_SYK} denoted "here", 
which is related to the chemical potential used in \cite{Complex_SYK-Large_q,Complex_SYK_Density_of_states,Tikhanovskaya_2021} and in our large $p$ computations in Section \ref{section_largepSYK}
and denoted "there"  by $\mu_{\textrm{here}}=-\beta\mu_{\textrm{there}}/2$. Using that $Q_0=\frac{1}{2}\tanh(\frac{\beta\mu_{\textrm{there}}}{2})$, we obtain from Eq.~(\ref{g_solution0}) 
\begin{align}
g(\tau) =-\frac{4\pi v}{\beta}\tan\left(\frac{\pi v}{2}\right)Q_0\left(\tau-\beta\Big(Q_0+\frac{1}{2}\Big)\right) +\log\left(\frac{\cos^2\left(\frac{\pi v}{2}\right)}{\cos^2\left(\frac{\pi v}{2}-\frac{\pi v\tau}{\beta}\right)}\right), 
\label{g_solution2}
\end{align} 
with 
\begin{align}
\beta\mathcal{J}=\frac{\pi v}{\cos\left(\frac{\pi v}{2}\right)}{\rm e}^{-2\pi v\tan\left(\frac{\pi v}{2}\right)Q_0^2},
\label{betaJ_solution2}
\end{align} 
where we identify $\mathcal{J}=\bold{J}\gamma(\mu)$.
Equations (\ref{g_solution2},\ref{betaJ_solution2}) agree with our result of large $p$ calculations, Eqs.~(\ref{g_solution},\ref{betaJ_solution}). 
Note solution $g(\tau)$ in Eq.~(\ref{g_solution2}) has antisymmetric and symmetric terms with respect to reflection around $\beta/2$. 
Our result also agrees with the large $p$-solution at $\mu=0$, symmetric part, in \cite{Large_q}.

\section{Bulk geometry and holography} 
\label{section_metric}

Ref.~\cite{SS10} pointed out the holographic connection between a model closely related to the complex SYK model and AdS$_2$ black hole horizons. This connection was at the level of the large-$N$ limit of the entropy and conformal correlators. Later, it was realized that there is a fully quantum connection at finite $N$ between the Majorana SYK model and Jackiw-Teitelboim gravity in 1+1 spacetime dimensions \cite{Kitaev_Suh,Large_q, Large_q_Schwarzian,CGPS}. This section will explore this connection for the complex SYK model, and show that an additional $U(1)$ gauge field is needed in the bulk \cite{Sachdev:2019bjn,Complex_SYK-Large_q,Complex_SYK_Density_of_states}. 

In order to find a bulk interpretation, which gives insight on the gravity dual description, we use an effective action for bi-local $G$ and $\Sigma$ fields. In the large $p$-limit
we consider fluctuations of the bi-local fields around the free solutions $G_0$ and $\Sigma=0$ \cite{Large_q_Liouville, Double_scaled_SYK_small_lambda}. 
Integrating out one of the fields we get an effective action for a bi-local Green's function fluctuation. 
We consider an average of two times as a time variable and the difference as a space variable. As a result we get an effective
theory defined on a two dimensional space. The goal is to extract a metric of this space and find a gravity dual description. This logic is similar to constructing a kinematic space with a volume form, which is used to express the length
of a curve in the bulk, thus mapping the volume form to a metric of the kinematic space \cite{Kinematic_space,Kinematic_space2,Kinematic_space3}. 

\subsection{Action for the complex SYK model}
\label{sec:AcSYK}

We start with an effective $G\Sigma$ action for the complex SYK model Eq.~(\ref{effective_action})
\begin{align}
-I[G,\Sigma] &= N\ln{\rm det}\left [\delta(\tau_1-\tau_2)(\partial_{\tau_1}-\mu)-\Sigma(\tau_1,\tau_2)\right]\nonumber\\
&-N \int d\tau_1 d\tau_2\left[ \Sigma(\tau_1,\tau_2)G(\tau_2,\tau_1)-\frac{J^2}{p} [G(\tau_1,\tau_2)]^{p/2}[G(\tau_2,\tau_1)]^{p/2} \right].
\label{effective_action2}
\end{align} 
In the double scaling limit of Section~\ref{section_DSSYK}, this effective theory becomes a bi-local theory in time $\tau$, i.e. a local theory on the kinematic space
labeled by pairs of time instances $(\tau_1,\tau_2)$. Following \cite{Large_q_Liouville, Double_scaled_SYK_small_lambda}, we make an ansatz similar to that in the large $p$ limit of Section~\ref{section_largepSYK}
\begin{align}
G(\tau_1,\tau_2)=G_0(\tau_{12})\left(1+\frac{g(\tau_1,\tau_2)}{p}\right),\quad 
\Sigma(\tau_1,\tau_2)=\frac{\sigma(\tau_1,\tau_2)}{p},
\label{ansatz3}
\end{align} 
where $\tau_{12}=\tau_1-\tau_2$ and $G_0(\tau)=\frac{{\rm e}^{\mu\tau}}{1+{\rm e}^{\mu\beta}}$ for the finite temperature time $0< \tau< \beta$. We switch to the frequency space
\begin{align}
G(i\omega_n)=\int_0^{\beta} d\tau G(\tau){\rm e}^{i\omega_n \tau},\quad 
G(\tau)=\frac{1}{\beta}\sum_n G(i\omega_n){\rm e}^{-i\omega_n\tau}, 
\end{align} 
where $\omega_{n}=\frac{2\pi}{\beta}(n+\frac{1}{2})$ and the free Green's function is $G_0(\omega_n)=\frac{1}{-i\omega_n-\mu}$. Terms in the action Eq.~(\ref{effective_action2}) 
except for the last interaction term are given in the frequency space
\begin{align}
N\sum_{\omega}\left[ \log\left[-i\omega-\mu-\Sigma(\omega)\right]-\Sigma(\omega)G(\omega)\right].
\end{align} 
Using the large-$p$ ansatz Eq.~(\ref{ansatz3}) and expanding to the second order we have
\begin{align}
-\frac{N}{2p^2}\sum_{\omega_1,\omega_2}\frac{\sigma(\omega_1)}{-i\omega_1-\mu}\frac{\sigma(\omega_2)}{-i\omega_2-\mu}
-\frac{N}{2p^2}\sum_{\omega}2\sigma(\omega)\left[G_0\times g\right](\omega),
\end{align} 
where the linear term proportional to $\frac{1}{p}\frac{\sigma(\omega)}{-i\omega-\mu}$ cancels between the two terms. Taking the Gaussian integral over the $\sigma$ field in the action we obtain
\begin{align}
-\frac{N}{8p^2}\sum_{\omega_1,\omega_2}(-i\omega_1-\mu)\left[G_0\times g\right](\omega_1)\times(-i\omega_2-\mu)\left[G_0\times g\right](\omega_2)\,.
\label{action_frequency}
\end{align} 
Returning back to the time space in Eq.~(\ref{action_frequency}) and restoring the interaction term we have for the action 
Eq.~(\ref{effective_action2})
 \begin{align}
 -I=-\frac{N}{8p^2} \int d\tau_1d\tau_2(\partial_{\tau_1}-\mu)G_0(\tau_{12})g(\tau_1,\tau_2)\times (\partial_{\tau_2}-\mu)G_0(\tau_{21})g(\tau_2,\tau_1)&&\nonumber\\
 +\frac{N}{p}\int d\tau_1d\tau_2 J^2 [G_0(\tau_{12})]^{p/2}[G_0(\tau_{21})]^{p/2} {\rm e}^{g_s(\tau_1,\tau_2)}.&&
 \label{frequency_action}
\end{align} 

To confirm Eq.~(\ref{frequency_action}), we repeat these calculations in the time space. We consider $G_0(\tau)=(\partial_{\tau}-\mu)^{-1}$  to be an operator in the action Eq.~(\ref{effective_action2}).
Expanding the action in large $p$ and dropping an additive constant from the
determinant term and other constants, we have to the leading order
\begin{align}
-I = -\frac{N}{2p^2}\int d\tau_1 d\tau_2d\tau_3d\tau_4 G_0(\tau_{12})\sigma(\tau_2,\tau_3)G_0(\tau_{34})\sigma(\tau_4,\tau_1)&&\nonumber\\ 
-\frac{N}{2p^2} \int d\tau_1 d\tau_2 2 \sigma(\tau_1,\tau_2)G_0(\tau_{21})g(\tau_2,\tau_1)&&\nonumber\\
+N\int d\tau_1d\tau_2\frac{J^2}{p} [G_0(\tau_{12})]^{p/2}[G_0(\tau_{21})]^{p/2} {\rm e}^{g_s(\tau_1,\tau_2)}\,.&&
\label{effective_action33}
\end{align}
We can view the integrands as a matrix multiplication with times $\tau_i$ being indices. The free Green's function satisfies 
\begin{align}
(\partial_{\tau_1}-\mu)G_0(\tau_{12})=\delta(\tau_1-\tau_2)\,, 
\label{free_eq}
\end{align}
that becomes in the matrix form 
\begin{align}
\int d\tau_3 G_0^{-1}(\tau_{13})G_0(\tau_{32})=\delta(\tau_1-\tau_2)\,,
\label{}
\end{align} 
that gives 
\begin{align}
G_0^{-1}(\tau_{12})=\delta(\tau_1-\tau_2)(\partial_{\tau_1}-\mu).
\label{inverseG0}
\end{align} 
In Eq.~(\ref{effective_action33})
$g(\tau_1,\tau_2)$ has symmetric and antisymmetric terms with respect to permutation $\tau_1\leftrightarrow \tau_2$, with $g_s(\tau_1,\tau_2)=\frac{1}{2}(g(\tau_1,\tau_2)+g(\tau_2,\tau_1))$ and $g_a(\tau_1,\tau_2)=\frac{1}{2}(g(\tau_1,\tau_2)-g(\tau_2,\tau_1))$. Following \cite{2SYK_Large_q,Large_q2,2SYK_Large_q3} we assume that in the SYK regime a dominant $g$ is symmetric, leading to a dominant $\sigma$ being antisymmetric with respect to $\tau_1\leftrightarrow \tau_2$.
Performing the Gaussian integral over $\sigma$ we have
\begin{align}
-I = -\frac{N^2}{8p^2}\int d\tau_1 d\tau_2 d\tau_3d\tau_4 G_0^{-1}(\tau_{12})G_0(\tau_{23})g(\tau_2,\tau_3)G_0^{-1}(\tau_{34})G_0(\tau_{41})g(\tau_4,\tau_1)&&\nonumber\\
+\frac{N}{p}\int d\tau_1d\tau_2 J^2 [G_0(\tau_{12})]^{p/2}[G_0(\tau_{21})]^{p/2} {\rm e}^{g_s(\tau_1,\tau_2)}.&&
\label{effective_action34}
\end{align}
Using an inverse free Green's function Eq.~(\ref{inverseG0}) it gives the same action that we obtained in the frequency space Eq.~(\ref{frequency_action}). 

Acting by the inverse Green's function on $[G_0g]$ structures and using Eq.~(\ref{free_eq}), the first integral in Eq.~(\ref{effective_action34}) becomes
\begin{align}
-\frac{N^2}{8p^2}\int d\tau_1 d\tau_2 \left[\delta(\tau_1-\tau_2-\varepsilon)g(\tau_1,\tau_2)+G_0(\tau_{12})\partial_{\tau_1}g(\tau_1,\tau_2)\right]&&\nonumber\\
\times
\left[\delta(\tau_2-\tau_1-\varepsilon)g(\tau_2,\tau_1)+G_0(\tau_{21})\partial_{\tau_2}g(\tau_2,\tau_1)\right],&&
\label{}
\end{align}
where we introduce $\varepsilon\to 0$ to regularize a product of two delta functions. Terms with the delta functions contribute the boundary values of $g(0^{+})$ and $g(0^{-})=g(\beta^{-})$. 
Using the free Green's function Eq.~(\ref{G0function}) $G_0(\tau)=\frac{\rm e^{\mu\tau}}{1+{\rm e}^{\mu\beta}}=\frac{\rm e^{\mu(\tau-\frac{\beta}{2})}}{2\cosh\left(\frac{\mu\beta}{2}\right)}$, we have
\begin{align}
-G_0(\tau_{12})G_0(\tau_{21})=\frac{1}{\left[2{\rm e}^{\frac{\mu\beta}{2}}\cosh\left(\frac{\mu\beta}{2}\right)\right]^2}\equiv b^2,
\label{}
\end{align}
note that $G_0(0)=\frac{1}{2\rm{e}^{\frac{\mu\beta}{2}}\cosh\left(\frac{\mu\beta}{2}\right)}=b$.
Rescaling the coupling constant
\begin{align}
\frac{\mathcal{J}^2}{2p}=\frac{J^2}{\left[2{\rm e}^{\frac{\mu\beta}{2}}\cosh\left(\frac{\mu\beta}{2}\right)\right]^{p-2}}\,,
\label{}
\end{align}
we obtain the Liouville effective action for the Green's function fluctuation in the complex case
\begin{align}
I =&   \frac{N}{8p^2}b^2\int d\tau_1 d\tau_2\left(\phantom{\frac{1}{1}}\hspace{-0.35cm}\left[\delta(b\tau_{12}-\varepsilon)g(\tau_1,\tau_2)+ \partial_{\tau_1}g(\tau_1,\tau_2)\right]\right.&&\nonumber\\
&\left.\times\left[\delta(b\tau_{21}-\varepsilon)g(\tau_2,\tau_1)+\partial_{\tau_2}g(\tau_2,\tau_1)\right]
-4\mathcal{J}^2 {\rm e}^{g_s(\tau_1,\tau_2)}\phantom{\frac{1}{1}}\hspace{-0.3cm}\right),&&
\label{result_action}
\end{align}
where the product of two delta functions vanishes due to a regularization with $\varepsilon$. The effective action Eq.~(\ref{result_action}) differs from the one obtained for the Majorana fermions in Refs.~\cite{Large_q_Liouville,Double_scaled_SYK_small_lambda} by extra delta functions.

Computing the variation of Eq.~(\ref{result_action}), we find that the Green's function satisfies the equation of motion 
 \begin{align}
g^{\prime\prime}(\tau_1,\tau_2)=2\mathcal{J}^2 {\rm e}^{g_s(\tau_1,\tau_2)}+\delta^{\prime}(b\tau_{12})g(\tau_1,\tau_2)+\delta(b\tau_{12})g^{\prime}(\tau_1,\tau_2).
\label{eq_of_motion2}
\end{align}
It is the Liouville equation with the boundary conditions, which are the terms with $\delta$ and $\delta^{\prime}$ functions. Integrating them twice gives at the boundary a step-function for $g$ and a cusp for $g^{\prime}$, i.e. on the finite temperature circle 
 $(0^{+},\beta^{-})$ and identifying $0^{-}=\beta^{-}$ the Green's function is discontinuous $g(0)\neq g(\beta)$ and has a cusp in the derivative $g^{\prime}(0)\neq g^{\prime}(\beta)$ plus a linear in time term, and the solution is the one we obtained earlier in Eq.~(\ref{g_solution}).

Following \cite{Large_q_Liouville,Double_scaled_SYK_small_lambda}, we introduce space $z$ and time $t$ variables in the two dimensional kinematic space
\begin{align}
z=\tau_1-\tau_2, \quad t=\tau_1+\tau_2.
\label{}
\end{align}
Rewriting different combinations in Eq.~(\ref{result_action}), we have for the diagonal part 
\begin{align}
\partial_{\tau_1}g(\tau_1,\tau_2)\partial_{\tau_2}g(\tau_2,\tau_1)=\left[-\partial_zg(z)\partial_zg(-z)+\partial_tg(z)\partial_tg(-z)\right]&&\nonumber\\
+\left[\partial_zg(z)\partial_tg(-z)-\partial_zg(-z)\partial_tg(z)\right],&& 
\label{metric0}
\end{align}
and for the mixed part
\begin{align} 
&\delta(\tau_{12})g(\tau_1,\tau_2)\partial_{\tau_2}g(\tau_2,\tau_1)+\delta(\tau_{12})g(\tau_2,\tau_1)\partial_{\tau_1}g(\tau_1,\tau_2)&& \nonumber\\
&\qquad =\delta(z)\left[-g(z)\partial_zg(-z)+g(-z)\partial_zg(z)\right]+\delta(z)\partial_t\left[ g(z)g(-z)\right],&&
\label{boundary_term0}
\end{align}
where we suppressed the time variable $t$ in the argument.
When integrated over $z$,  the mixed terms in the second bracket of Eq.~(\ref{metric0}) vanish  by exchanging $z\to -z$ in the integral, meaning that the off-diagonal components of the bulk metric tensor are zero.
The terms with delta function in Eq.~(\ref{boundary_term0}) are nonzero because of discontinuity of the two-point function $g$ and its derivative at the boundaries. This could create a flux at the boundary $z=0$ of the bulk kinematic space.
As a result we get a two dimensional effective action of the Liouville form plus the boundary term (this is equivalent to Eq.~(\ref{result_action}), but now written in the kinematic space):
\begin{align}
I &=  I_{\rm bulk}^{kinematic}+I_{\rm boundary}^{kinematic}\nonumber\\
& =\frac{N}{8p^2}b^2\int dz dt \left[-\frac{1}{2}(\partial_{z}g(t,z))^2+\frac{1}{2}(\partial_{t}g(t,z))^2-2\mathcal{J}^2 e^{g_s(t,z)}\right. \label{effective_action3}\\
& ~~~~~\left.+\frac{1}{2}\delta(z)\left\{-g(z)\partial_zg(-z)
+g(-z)\partial_zg(z)+\partial_t\left[ g(z)g(-z)\right]\right\}\right].
\label{effective_action4}
\end{align}
we suppressed $t$ index in the argument of the boundary action.
In the kinetic term, the field $g=g_s+g_a$ has symmetric and antisymmetric parts, while it is symmetric part alone $g_s=\frac{1}{2}(g(z)+g(-z))$ that enters the interaction term $\mathcal{J}^2 e^{g_s}$. Equation of motion from the bulk action Eq.~(\ref{effective_action3}) is
\begin{align}
\left(-\partial^2_t+\partial^2_z\right)g(t,z)=2\mathcal{J}^2 e^{g_s(t,z)}.
\label{equation_of_motion}
\end{align}
Keeping only $z$ dependence Eq.~(\ref{equation_of_motion}) splits in two equations for the symmetric and antisymmetric parts 
\begin{align}
\partial^2_zg_s(z)=2\mathcal{J}^2 e^{g_s(z)},\, \quad 
\partial^2_zg_a(z)=0,
\label{effective_theory_EOM}
\end{align}
which is Eq.~(\ref{eqs_of_motion}) obtained for the complex SYK.
Varying the boundary action Eq.~(\ref{effective_action4}) we obtain
nonzero boundary conditions $g_s(0)=g_s(\beta)\neq 0$ and $g_a(0)=-g_a(\beta)\neq 0$ which supplement equations of motion.
We recover equations of motion for symmetric Eq.~(\ref{symmetric})
and antisymmetric Eq.~(\ref{antisymmetric}) Green's functions with solution given by Eq.~(\ref{g_solution}).

We split the Green's function solution Eq.~(\ref{g_solution})
into symmetric $g_s$ and antisymmetric $g_a$ parts with respect to $z\to (\beta-z)$ reflection, with $g=g_s+g_a$,
and also separate symmetric part $g_s$ into a $z$-dependent $g_s^{(1)}$ and a constant $g_s^{(2)}$ contributions
\begin{align}
g_s &=g_s^{(1)}+g_s^{(2)} \,,
g_s^{(1)}(z) = 
\log\left(\frac{\cos^2\left(\frac{\pi v}{2}\right)}{\cos^2\left(\frac{\pi v}{2}-\frac{\pi v z}{\beta}\right)}\right),\,\,g_s^{(2)}= 4\pi v\tan\left(\frac{\pi v}{2}\right)Q_0^2,\label{effective_fields0}\\
g_a(z) &= -\frac{4\pi v}{\beta}\tan\left(\frac{\pi v}{2}\right)Q_0(z-\frac{\beta}{2}),
\label{effective_fields}
\end{align}
where equation (\ref{betaJ_solution}) for the coupling constant can be written 
\begin{align}
\mathcal{J}^2=\left(\frac{\pi v}{\beta}\right)^2\frac{e^{-g_s^{(2)}}}{\cos^2\left(\frac{\pi v}{2}\right)},
\label{beta_J2}
\end{align}
which shows that an additional constant solution $g_s^{(2)}$ in the charged SYK amounts to rescaling (renormalization) of the coupling constant $\mathcal{J}$. 

\subsection{2D dilaton gravity}


In this subsection
we construct a correspondence between the effective theory for the Greens function correction $g(z,t)$ Eq.~(\ref{effective_action3},\ref{effective_action4}) in the complex SYK, and a charged version of a specific 2D dilaton gravity.

There exist many 2D dilaton gravity models that admit an AdS ground state \cite{2dGravity,2dGravity2}. Hartman and Strominger  \cite{Maxwell_Dilaton0} suggested an approach in the context of 2D Maxwell-dilaton gravity theories with an $\textrm{AdS}_2$ ground state supporting a constant electric field.
We add to the Jackiw-Teitelboim model  a minimally coupled $U(1)$ gauge field with the bulk action \cite{2d_Maxwell_Dilaton}
\begin{align}
 I_{\rm bulk}^{\rm gravity} \sim \int d^2x\sqrt{-G}\left[e^{-2\Phi}\left(R+2\mathcal{J}^2\right)-\frac{1}{\mathcal{J}^2}F^2\right],
\label{2d_gravity}
\end{align}
where $G$ is the determinant of the metric tensor taken with the lower indices, 
$\Phi$ is the dilaton field, $F^2=F_{\mu\nu}F^{\mu\nu}$ is the square of the field strength, and the relation between the field strengths with upper and lower indices is
$F^{\mu\nu}=G^{\mu\gamma}F_{\gamma\delta}G^{\delta\nu}$ and $G^{\mu\gamma}F_{\gamma\delta}=F^{\mu}_{\delta}$. For the field strength in the curved space-time, covariant derivative is reduced to a simple derivative 
$F_{\mu\nu}=\nabla_{\mu}A_{\nu}-\nabla_{\nu}A_{\mu}
=\partial_{\mu}A_{\nu}-A_{\tau}\Gamma^{\tau}_{\mu\nu}-\partial_{\nu}A_{\mu}+A_{\tau}\Gamma^{\tau}_{\nu\mu}=\partial_{\mu}A_{\nu}-\partial_{\nu}A_{\mu}$, since Christoffel symbols are symmetric $\Gamma^{\tau}_{\mu\nu}=\Gamma^{\tau}_{\nu\mu}$ in the curved space-time without torsion. 


We establish a matching between the fields of the effective (kinematic) Eq.~(\ref{effective_action3},\ref{effective_action4}) and gravity Eq.~(\ref{2d_gravity}) theories in the bulk.
As we solve equations of motion we obtain explicit relations between the field content of the two theories.
Taking variation of the action Eq.~(\ref{2d_gravity}) with respect to the fields we get equations of motion
\begin{align}
&\delta G^{\mu\nu}:\,\,
(\nabla_{\mu}\nabla_{\nu}-G_{\mu\nu}\nabla^2)e^{-2\Phi}+G_{\mu\nu}\mathcal{J}^2e^{-2\Phi}+\frac{2}{\mathcal{J}^2}(F^{\lambda}_{\mu}F_{\nu\lambda}-\frac{1}{4}G_{\mu\nu}F^2)=0,&\label{eom1}\\
&\delta\Phi:\,\,
-2e^{-2\Phi}\left(R+2\mathcal{J}^2\right)=0,&\label{eom2}\\
&\delta A_{\mu}:\,\, 
\frac{4}{\mathcal{J}^2}\nabla^{\nu}F_{\nu\mu}=0,&\label{eom3}
\end{align}
where in Eq.~(\ref{eom1}) we used the Jacobi formula $d \det A=\det A \cdot tr(A^{-1}dA)$ that gives $\delta \sqrt{-G}=-\frac{1}{2}\sqrt{-G}\left(G_{\mu\nu}\delta G^{\mu\nu}\right)$,
and the covariant derivative in Eq.~(\ref{eom3}) is given by
$\nabla_{\alpha}F^{\alpha\beta}=\partial_{\alpha}F^{\alpha\beta} +\Gamma_{\mu\alpha}^{\alpha}F^{\mu\beta}+\Gamma_{\mu\alpha}^{\beta}F^{\alpha\mu}$.
When contracted with $G^{\mu\nu}$, eq. (\ref{eom1}) reduces to
\begin{align}
(-\nabla^2+2\mathcal{J}^2)e^{-2\Phi}+\frac{1}{\mathcal{J}^2}F^2=0.
\label{eom4}
\end{align}
One can find classical solutions of Eqs.~(\ref{eom1}-\ref{eom4}) in a closed form.
Solution with a constant dilaton exhibit an enhanced symmetry. Indeed  classical solutions of Eq.~(\ref{eom2}) must be space-times of constant (negative) curvature. This space is maximally symmetric and exhibits three Killing vectors, meaning that it is locally and asymptotically $AdS_2$. 
Solutions with a non-constant dilaton breaks the $SL(2,R)$ algebra generated by
the Killing vectors to $U(1)$.
In the regimes of the double-scaled SYK and the Jackiw-Teitelboim gravity full $AdS_2$ algebra is broken \cite{Double_scaled_SYK,Large_q_Schwarzian}. Therefore we write
Eqs.~(\ref{eom1}-\ref{eom4}) for a non-constant dilaton 
\begin{align}
&&\left(-\Delta +2\mathcal{J}^2\right)
e^{-2\Phi}+\frac{1}{\mathcal{J}^2}F^2=0,\label{eom_constant1}\\
&& R+2\mathcal{J}^2=0,\label{eom_constant2}\\
&&\nabla^{\nu}F_{\nu\mu}=0,
\label{eom_constant3}
\end{align}
where $\Delta=\nabla^2$, $\Delta f=\frac{1}{\sqrt{-G}}\partial_i\left(\sqrt{-G}G^{ij}\partial_j f\right)$ is the Laplace operator in the curved space-time.
Eq.~(\ref{eom_constant2}) gives the scalar curvature
\begin{align}
R=-2\mathcal{J}^2,
\label{Ricci_scalar}
\end{align}
that is valid in any coordinate system.
Equation (\ref{eom_constant3}) is solved by the covariantly constant field strength which can be written through a constant  electric field $E$
\begin{align}
F_{\mu\nu}=2E\,\varepsilon_{\mu\nu},
\label{field_strength2}
\end{align}
where the Levi-Civita tensor in the curved space-time is defined as 
$\varepsilon_{12}=-\varepsilon_{21}=\sqrt{-G}$, therefore
contraction with the metric tensor is $G^{\mu\gamma}\varepsilon_{\gamma\delta}G^{\delta\nu}\varepsilon_{\mu\nu}=-2$. Using contraction with the Levi-Civita tensor we have $F^2=-8E^2$, equation (\ref{eom_constant1}) becomes
\begin{align}
\left(-\Delta+2\mathcal{J}^2\right)
e^{-2\Phi}-\frac{8}{\mathcal{J}^2}E^2=0.
\label{connection_phi_e}
\end{align}

\subsubsection{Metric and Ricci scalar}

In what follows we choose the coordinate system and fix the $U(1)$ gauge $A_z=0$,
\begin{align}
ds^2=G_{zz}(z,t)dz^2+G_{tt}(z,t)dt^2,\,\, A_{\mu}dx^{\mu}=A_t(z,t)dt,
\label{coordinate_system}
\end{align}
Quantizing gravity in two dimensions Eq.~(\ref{2d_gravity}), Liouville theory describes the quantum mechanics of the Weyl factor
with respect to the naive metric appearing in the kinetic term. We make an ansatz for the Lorentzian metric 
\begin{align}
G_{\mu\nu}=e^{g_s}\eta_{\mu\nu}, 
\label{AdSmetric2}
\end{align}
where $\eta_{\mu\nu}=(-1,1)$ in a flat space-time, 
$G_{\mu\nu}G^{\mu\nu}=2$ and $\mu,\nu=1,2$. Explicitly the metric with lower and upper indices (the first entry $tt$, the second one $zz$) is 
$G_{\mu\nu}={\rm diag}\left(-e^{g_s},e^{g_s}\right)$, 
$G^{\mu\nu}={\rm diag}\left(-e^{-g_s},e^{-g_s}\right)$.
In the bulk gravity action Eq.~(\ref{2d_gravity}) $G$ is a determinant of the metric tensor taken with lower indices, therefore a square root of the negative determinant is $\sqrt{-G}={\rm e}^{g_s}$.

Using the metric tensor Eq.~(\ref{AdSmetric2}) 
Christoffel symbols are $\Gamma^{z}_{zz}=\Gamma^{z}_{tt}=\Gamma^{t}_{zt}=\Gamma^{t}_{tz}=\frac{1}{2}\partial_zg_s$, the Ricci curvature  tensor components are $R_{tt}=-R_{zz}=\frac{1}{2}\partial^2_zg_s$, and the Ricci scalar curvature $R=G^{\mu\nu}R_{\mu\nu}$ becomes 
\cite{Large_q,Double_scaled_SYK_small_lambda,2dGravity}
\begin{align}
R=-e^{-g_s}\partial^2_zg_s=\Delta g_s,
\label{curvature22}
\end{align}
where $g_s=\frac{\delta G}{G_0}$. Here
the Laplace operator in the curved space-time is given by
$\Delta f =\frac{1}{\sqrt{-G}}\partial_i\left(\sqrt{-G}G^{ij}\partial_j f\right)$, that with the metric tensor Eq.~(\ref{AdSmetric2}) becomes 
\begin{align}
\Delta f=e^{-g_s}(-\partial_t^2+\partial_z^2)f.
\label{Laplace}
\end{align}
Note that, at low temperatures, for $v\to 1$, the Casimir operator of $SL(2,R)$ group acting on the two times $\tau_1,\tau_2$ is \cite{Large_q,Large_q2}
\begin{align}
C_{12}\Psi=-(\tau_1-\tau_2)\partial_{\tau_1}\partial_{\tau_2}\Psi=z^2(-\partial^2_{t}+\partial^2_{z})\Psi=\Delta_{{\mathcal J}=1} \Psi,
\label{Casimir}
\end{align}
where $\Psi=\frac{\delta G}{G_c}$, with $\delta G$ is the small fluctuation around classical solution which is the conformal Green's function $G_c$ in the IR. Therefore Casimir operator is 
$\Delta_{{\mathcal J}=1}$ which is the Laplacian in the $AdS_2$     
in coordinates $ds^2=\frac{-dt^2+dz^2}{z^2}$ Eq.~(\ref{gs_IR}) where the AdS radius is set to one. It turns out that Casimir commutes with the kernel for the four point function, meaning that they have the same set of eigenfunctions. It was found in \cite{Large_q} that eigenfunctions of Casimir $C_{1+2}$ were particular hypergeometric functions related to conformal blocks of weight h, $C=h(h-1)$. In what follows we obtain hypergeometric function for the dilaton solution. It shows that there is a connection between SL(2) algebra of the field theory and the solutions in the gravity theory.    

Using Eq.~(\ref{curvature22}) $R=-e^{-g_s}\partial^2_zg_s$, the Ricci curvature condition 
Eq.~(\ref{Ricci_scalar}) can be written in the form of a Liouville's differential equation 
\begin{align}
\partial_z^2g_s=2\mathcal{J}^2e^{g_s},
\label{gravity_theory_EOM}
\end{align}
with the solution $g_s$ given by Eq.~(\ref{g_solution},\ref{gs},\ref{gs22}) and Eq.~(\ref{effective_fields0}). Therefore when the metric factor is given by Eq.~(\ref{AdSmetric2}) the metric equation 
(\ref{eom_constant2}) is reduced to an equation of motion for the symmetric Green's function Eq.~(\ref{symmetric},\ref{effective_theory_EOM}). 

Substituting solution for the symmetric part $g_s(z)$ Eq.~(\ref{effective_fields0}) into the metric factor Eq.~(\ref{AdSmetric2}) we have  
\begin{align}
{\rm e}^{g_s(z)}=\frac{\cos^2\left(\frac{\pi v}{2}\right)}{\cos^2\left(\frac{\pi v}{2}(1-\frac{2z}{\beta})\right)}\, e^{4\pi v \tan\left(\frac{\pi v}{2}\right)Q_0^2}\,.
\label{gs}
\end{align}
Using the relation between $v$ and $\beta\mathcal{J}$ Eq.~(\ref{beta_J2})  
\begin{align}
\beta\mathcal{J}=\frac{\pi v}{\cos\left(\frac{\pi v}{2}\right)}
e^{-2\pi v\tan\left(\frac{\pi v}{2}\right)Q_0^2}\,,
\label{betaJ_solution22}
\end{align}
we can rewrite the metric factor as
\begin{align}
e^{g_s(z)}=\left(\frac{\pi v}{\beta}\right)^2\frac{1}{\mathcal{J}^2\cos^2\left(\frac{\pi v}{2}(1-\frac{2z}{\beta})\right)}.
\label{gs22}
\end{align}
In the IR region at strong coupling $\beta{\mathcal J}\gg 1$, corresponding to $v\to 1$,  the metric is
\begin{align}
e^{g_s}=\frac{1}{{\mathcal J}^2z^2}\,,
\label{gs_IR}
\end{align}
and relating to the notations of the previous section $\mathcal{J}=\bold{J}\gamma(\mu)$. Therefore in the conformal limit $\beta{\mathcal J}\gg 1$ the metric Eq.~(\ref{AdSmetric2},\ref{gs_IR}) describes a two-dimensional $AdS_2$ with radius given by $a=1/{\mathcal J}$. 

Let us examine the curvature of the kinematic space on the field theory side.
Representing the bulk effective action Eq.~(\ref{effective_action3}) as
\begin{align}
I_{\rm bulk}^{kinematic}\sim \int dzdt\sqrt{-G}\left[\frac{1}{2}G^{\mu\nu}\partial_{\mu}g\partial_{\nu}g-2\mathcal{J}^2 \right],
\label{2d_effective}
\end{align}
where $\sqrt{-G}=e^{g_s}$ and the metric tensor is given by Eq.~(\ref{AdSmetric2}),
we obtain the Ricci scalar $R=-e^{-g_s}\partial^2_zg_s$ Eq.~(\ref{curvature22}).
Using the second derivative of the symmetric part $g_s$ Eq.~(\ref{gs22})  
$\partial^2_z g_s=2\left(\frac{\pi v}{\beta}\right)^2\sec^2\left(\frac{\pi v}{2}(1-\frac{2z}{\beta})\right)$ and the relation Eq.~(\ref{betaJ_solution22}), the Ricci scalar Eq.~(\ref{curvature22}) equals to
$R=-2\mathcal{J}^2$. 
Thus there is an equivalence between the effective (kinematic) field theory Eq.~(\ref{effective_action3},\ref{2d_effective}) and the gravity theory Eq.~(\ref{2d_gravity}) for the equations of motion and for calculated variables, e.g. the Ricci curvature scalar Eq.~(\ref{Ricci_scalar}). 

We have obtained a negative constant  scalar curvature Eq.~(\ref{Ricci_scalar}), meaning that it is a two-dimensional Anti-de Sitter space-time  with  corresponding radius $a$ equals to
\begin{align}
a=\frac{1}{\mathcal{J}}\,.
\label{radius}
\end{align}
Note that the scalar curvature of the bulk $AdS_2$ spacetime Eq.~(\ref{Ricci_scalar}) is the same for the complex SYK and for the Majorana fermions \cite{Double_scaled_SYK_small_lambda}. 
Writing the metric through a line element 
\begin{align}
ds^2 ={\rm e}^{g_s(z)}\left(-dt^2 +dz^2\right),
\label{metric}
\end{align}
it has a form of the $AdS_2$ spacetime upon imposing the Liouville equations of motion (\ref{equation_gs}), (\ref{eq_of_motion2}) for the symmetric field $g_s$ with solution given by Eq.~(\ref{gs},\ref{gs22}). 
At low temperatures, corresponding to $v\rightarrow 1$, the metric factor Eq.~(\ref{gs22}) becomes
\begin{align}
ds^2=\left(\frac{\pi v}{\beta}\right)^2\frac{\left(-dt^2 +dz^2\right)}{\mathcal{J}^2\sin^2\left(\frac{\pi z}{\beta}\right)},
\label{}
\end{align}
which is Lorentzian patch of the ${\rm AdS}_2$ geometry, because $z$ is fixed and only part of the space is covered. Rescaling $z$ and $t$ as follows
$\sigma=\frac{\pi z}{\beta}$ and $\tau=\frac{\pi t}{\beta}$, we obtain
\begin{align}
ds^2=\frac{\left(-d\tau^2 +d\sigma^2\right)}{\mathcal{J}^2\sin^2\left(\sigma\right)},
\label{metric_zeroT}
\end{align}
which is a global Lorentzian ${\rm AdS}_2$ metric where $0<\sigma<\pi$.
For the finite temperature, when $0<v<1$, the metric is given
\begin{align}
ds^2=\left(\frac{\pi v}{\beta}\right)^2\frac{\left(-dt^2 +dz^2\right)}{\mathcal{J}^2\cos^2\left(\frac{\pi v}{2}(1-\frac{2z}{\beta})\right)},
\label{metric_finiteT}
\end{align}
which is a patch in the ${\rm AdS}_2$. 
We obtain the same Ricci scalar $R=-2\mathcal{J}^2$ for the zero Eq.~(\ref{metric_zeroT}) 
and finite Eq.~(\ref{metric_finiteT}) temperatures, confirming that it is a two-dimensional Anti-de Sitter space-time. 

\subsubsection{Maxwell field}

In Ref.~\cite{Sachdev:2019bjn}, the bulk electric field, $E$, was identified with the time  derivative of a phase field $\phi$ on the boundary, and the co-efficient of the bulk $E^2$ Maxwell term in the action was linked to the compressibility of the boundary theory. In the context of complex SYK, Refs.~\cite{Complex_SYK-Large_q,Complex_SYK_Density_of_states}, $\phi$ was linked to the phase of the fermion Green's function, and the compressibility was the coefficient of $(\partial_\tau \phi)^2$. In the present formulation, $\phi$ is linearly related to $g_a$, and so we will therefore connect the time derivative of $g_a$ to the bulk electric field $E$ in Eq.~(\ref{electric_field}).

Equation of motion (\ref{eom_constant3}) with identification Eq.~(\ref{field_strength2}) is written in the coordinate system Eq.~(\ref{coordinate_system})
with the metric
Eq.~(\ref{AdSmetric2})
\begin{align}
\partial_z E=0,
\end{align}
that 
gives a constant electric field $E={\rm const}$.
Making the following identification
\begin{align}
E=\partial_z g_a = -\frac{4\pi v}{\beta}\tan\left(\frac{\pi v}{2}\right)Q_0, 
\label{electric_field}
\end{align}
we satisfy equation for $g_a$
\begin{align}
\partial_z^2 g_a=0,
\end{align}
with the solution given by Eq.~(\ref{g_solution}) and Eq.~(\ref{effective_fields}). Therefore in the coordinate system Eq.~(\ref{AdSmetric2}) the equation (\ref{eom_constant3}) for the Maxwell field 
reduces to an equation of motion for the antisymmetric Green's function Eq.~(\ref{antisymmetric},\ref{effective_theory_EOM}). 
Expression for the field strength $F_{\mu\nu}=2E\,\varepsilon_{\mu\nu}$ Eq.~(\ref{field_strength2}) with the Levi-Civita tensor is
$\varepsilon_{12}=-\varepsilon_{21}=\sqrt{-G}=e^{g_s}$ gives an
equation for the vector potential 
\begin{align}
-\partial_zA_t=2E\sqrt{-G}=2Ee^{g_s}.
\label{A_t}
\end{align}
Using an equation of motion for the symmetric Green's function (\ref{effective_theory_EOM},\ref{gravity_theory_EOM}), we integrate Eq.~(\ref{A_t}) and find the vector potential
\begin{align}
A_t=\frac{2E}{\mathcal{J}^2}\partial_zg_s(0),
\label{A_t2}
\end{align}
where the electric field $E$ is given by Eq.~(\ref{electric_field}).
As expected the electric field and gauge potential vanish in the neutral case $Q_0=0$.
Solution for the metric and the gauge field potential in Eq.~(\ref{coordinate_system}) can be simplified by fixing the residual gauge freedom by $U(1)$ transformation
$A_{\mu}\to A_{\mu}+\partial_{\mu} \lambda$. 

\subsubsection{Dilaton field}

Using Laplace operator in the curved space-time
$\Delta f=\frac{1}{\sqrt{-G}}\partial_i\left(\sqrt{-G}G^{ij}\partial_jf\right)$ which in the coordinate system
Eq.~(\ref{AdSmetric2})
is $\Delta f=-e^{-g_s}\partial_z^2f$,
the equation of motion (\ref{connection_phi_e})
for the dilaton is written   
\begin{align}
\left(e^{-g_s}\partial_z^2+2\mathcal{J}^2\right)
e^{-2\Phi}=\frac{8}{\mathcal{J}^2}E^2.
\label{connection_phi_e2}
\end{align}
Using equation (\ref{gs22}) for the metric tensor 
we rewrite equation of motion (\ref{connection_phi_e2}) for the dilaton 
\begin{align}
\left[\left(\frac{\beta}{\pi v}\right)^2\cos^2\left(\frac{\pi v}{2}-\frac{\pi v z}{\beta}\right)\partial_z^2 +2\right]e^{-2\Phi}=\frac{8}{\mathcal{J}^4}E^2.
\label{connection_phi_e23}
\end{align}
In order to calculate the r.h.s. of Eq.~(\ref{connection_phi_e23}) we use equation (\ref{betaJ_solution22}) for 
$\beta\mathcal{J}$ 
\begin{align}
\mathcal{J}^2=\left(\frac{\pi v}{\beta}\right)^2\frac{e^{-g_s^{(2)}}}{\cos^2\left(\frac{\pi v}{2}\right)}\,,
\label{}
\end{align}
with $g_s^{(2)}=4\pi v \tan\left(\frac{\pi v}{2}\right)Q_0^2$ given in Eq.~(\ref{effective_fields0}),
and the electric field $E$ given by Eq.~(\ref{electric_field}).
Rescaling the variable
$\frac{\pi v}{2}-\frac{\pi v z}{\beta}= \tilde{z}$ we get an equation for $y(\tilde{z})=e^{-2\Phi}$ 
\begin{align}
\cos^2(\tilde{z}) \partial_{\tilde{z}}^2y +2y=b_e,
\label{}
\end{align} 
where we introduced a constant b-electric
$b_e=8\left(\frac{\beta}{\pi v}\right)^4\cos^4\left(\frac{\pi v}{2}\right)e^{2g_s^{(2)}}E^2$, which is nonzero in the charged case.
Solution is given by the hypergeometric function
\begin{align}
&&y(\tilde{z})=\frac{b_e}{2}+c_1(-1)^{\frac{1}{4}(1-i\sqrt{7})}\cos^{\frac{1}{2}(1-i\sqrt{7})}(\tilde{z})
{}_2F_1\left(\frac{1}{4}-\frac{i\sqrt{7}}{4},\frac{1}{4}-\frac{i\sqrt{7}}{4};1-\frac{i\sqrt{7}}{2};\cos^2(\tilde{z})\right)\nonumber\\
&&+c_2(-1)^{\frac{1}{4}(1+i\sqrt{7})}\cos^{\frac{1}{2}(1+i\sqrt{7})}(\tilde{z})
{}_2F_1\left(\frac{1}{4}+\frac{i\sqrt{7}}{4},\frac{1}{4}+\frac{i\sqrt{7}}{4};1+\frac{i\sqrt{7}}{2};\cos^2(\tilde{z})\right)\,,
\label{hypergeometric}
\end{align} 
where $\tilde{z}=\frac{\pi v}{2}-\frac{\pi v z}{\beta}$
and constants $c_1,c_2$ can be found from the boundary conditions for the dilaton. Dilaton exists for the neutral Majorana case when $b_e=0$.
At the singularities $\tilde{z}=\frac{\pi}{2}+\pi n,\, n\in \mathbb{Z}$ (see Appendix \ref{AppGeodEq}), the dilaton field is not defined due to a prefactor which is a zero raised to an imaginary power.
At low temperatures, corresponding to $v\to 1$, or the conformal limit $\beta{\mathcal J}\gg 1$ the metric is a global Lorentzian AdS$_2$ geometry $ds^2=\frac{-d\tau^2+d\sigma^2}{{\mathcal J}^2\sin^2(\sigma)}$ Eq.~(\ref{metric_zeroT}) where space and time varaibles $z,t$ are rescaled $\sigma=\frac{\pi z}{\beta}$,  
$\tau=\frac{\pi t}{\beta}$ and $0< \sigma < \pi$. In the conformal limit the dilaton solution $y=e^{-2\Phi}$ Eq.~(\ref{hypergeometric}) diverges at the boundary $z=0$ ($\sigma=0$). 
This agrees with
the neutral dilaton solution in \cite{Large_q_Schwarzian} where the dilaton is diverging near the boundary and a new dimensionful coupling
constant is introduced which is the strength of that divergence.
The solution breaks the $SL(2)$ isometries to $U(1)$. 

We can represent the bulk effective action in the kinematic space  Eq.~(\ref{effective_action3},\ref{2d_effective}) through a kernel as
$\int d^2x\left[ g\left(\tilde{K}^{-1}-1\right)g\right]$ where a (symmetric) kernel $\tilde{K}$
generates a set of ladder diagrams for the four point function \cite{Large_q}. The effective action in the kinematic space Eq.~(\ref{effective_action3}) is mapped to the bulk gravity action Eq.~(\ref{2d_gravity}), which can be written (without an electric field) as $\int d^2x e^{-2\Phi}\left( -\partial_z^2g_s+2{\mathcal J}^2e^{g_s}\right)$ or $\int d^2xe^{g_s}e^{-2\Phi}\left(\Delta g_s+2{\mathcal J}^2\right)$
using the scalar curvature $R$ given in Eq.~(\ref{curvature22}). The eigenvectors of the kernel $\tilde{K}\Psi=\kappa\Psi$ and the eigenfunctions of Casimir operator Eq.~(\ref{Casimir}), which commutes with the kernel, are linear combinations of particular hypergeometric functions ${}_2F_1(\dots)$. These eigenfuctions were obtained in the conformal limit $\beta{\mathcal J}\gg 1$ in \cite{Large_q}.
It is interesting that the dilaton solution $e^{-2\Phi}$ Eq.~(\ref{hypergeometric}) is expressed through the hypergeometric functions signaling duality between the field theory and gravity actions and for certain matrix elements. 

We obtained the following mapping between the fields of the effective Eq.~(\ref{effective_action3}) and gravitational Eq.~(\ref{2d_gravity}) theories
\begin{align}
G_{\mu\nu}=(-e^{g_s},e^{g_s}),\,\,
e^{-2\Phi}=y\left(\frac{\pi v}{2}-\frac{\pi v z}{\beta}\right),\,\,
E=\partial_zg_a,\,\,
A_t=\frac{2E}{\mathcal{J}^2}\,\partial_zg_s(0),
\label{mapping3}
\end{align}
with the Greens functions $g_s, g_a$ are given by Eq.~(\ref{effective_fields0},\ref{effective_fields}) and solution $y$ is given by Eq.~(\ref{hypergeometric}). 
Dilaton exists in the $AdS_2$ space-time in a neutral case while an electric field vanishes at $Q_0=0$, providing gravity dual description for the SYK model with Majorana fermions.
Effective and gravitational theories are 
equivalent to each other on the level of equations of motion and they give the same calculated observables.

\section{Concluding Remarks}

In this paper we studied the complex SYK model beyond the conformal symmetry regime, where the problem becomes analytically tractable in the limit of large $p$. 
Given our lengthy analysis, we identify the key results of our paper in this section. 

Section~\ref{section_largepSYK} examined the complex SYK model in the limit of large $N$, followed by the limit of large $p$. Our result for the fermion Green's function is in Eq.~(\ref{g_solution}), and that for the grand potential is in Eq.~(\ref{Omega}). As compared to the SYK model with Majorana fermions the Green's function in the complex SYK case with nonzero chemical potential is not symmetric with respect to $\beta/2$ but is tilted due to different nonzero boundary values. We obtain extra antisymmetric linear in time  $g_a(\tau)\sim Q_0(\tau-\beta/2)$ and symmetric constant $g_s(0)\sim Q_0^2\beta$ terms in the Green's function that depend on the free fermions charge density $Q_0$. In the limit we are working, the partition function depends only on the symmetric part of the Green's function. As a result the grand potential Eq.~(\ref{Omega}) and the energy Eq.~(\ref{FreeEnergy}) are modified by the $Q_0^2$ terms accordingly. At strong coupling we find the solution for the Green's function in Eq.~(\ref{g_solution}) only at small chemical potentials, namely we must tune $\mu$ down as $\beta{\mathcal J}$ grows (in the Majorana SYK, or complex SYK with $\mu=0$,  the solution exists for all $\beta{\mathcal J}$).
This translates in the range for the charge density of free fermions $Q_0\leq 1/(2e\beta{\mathcal J})$ and for the parameter $v<v_{*}=1-4Q_0^2$ which parametrizes $\beta{\mathcal J}$ in the solution Eq.~(\ref{g_solution}). In the double-scaling limit $v$ is related to the energy at the saddle point. This means  that only small energies are allowed. Restrictions in the parameter range for $\mu<\mu_{*}$ and $v<v_{*}$ may be related to the complex SYK model undergoing a first-order phase transition at some critical $\mu=\mu_{*}$ and fixed 
$\beta{\mathcal J}$. It is possible that in order to avoid these difficulties we should use another ansatz to find Green's function at large $p$\footnote{We thank  Vladimir Narovlansky for discussions on this point}. We find that the large $p$ Green's function Eq.~(\ref{g_solution}) agrees with the conformal solution for the Green's function in complex SYK at leading order in both large $p$ and large 
$\beta{\mathcal J}$.       

Section~\ref{section_DSSYK} studied the double-scaling limit of large $N$, $p$ with $\lambda = p^2/N$ fixed.  We studied the small $\lambda$ limit without restricting to low energies as we did in the large $p$ limit SYK in Section ~\ref{section_largepSYK}.  We used results of Berkooz {\it et al.} \cite{Double_scaled_complex_SYK} for double-scaled complex SYK, which were expressed in terms of infinite series of moments of observable operators. We showed that these series could be resumed in the limit of small $\lambda$, and the results were in precise agreement with those of Section~\ref{section_largepSYK}. The result for the grand potential in Eq.~(\ref{free_energy}) agrees with that in Eq.~(\ref{Omega}), while the result for the Green's function in Eqs.~(\ref{g_solution2},\ref{betaJ_solution2}) agrees with that in Eqs.~(\ref{g_solution},\ref{betaJ_solution}). 
Operators in double-scaling SYK are characterized by the number of fermions that the operators are comprised of. Considering the two-point function we introduced $\lambda$ parameters associated with the sizes of the Hamiltonian and the operator for which we calculated the Green's function (number of (anti)fermions) as well as with the total fermion charge (difference between number of fermions and antifermions). We evaluated integrals in the classical limit of double-scaling SYK where $\lambda\to 0$ associated with size of the Hamiltonian, keeping only the leading order in the $1/\lambda$ expansion. This gave saddle point equations for general operators, including heavy operators. Since we are interested in the bulk geometry, we considered probe operators which are not too heavy in order not to backreact strongly on the geometry. We solved saddle point equations and extracted the result for the Green's function in the limit of small $\lambda$'s which are related to the size of the propagating operator and the charge density. To the leading order, solution of the saddle point equations was the same as in the neutral SYK.
Additional terms in the Green's function compared to a neutral case arose from $B$ factors Eqs.~(\ref{coefficients},\ref{AB}), that entered exponents in the moment for the two-point function Eq.(\ref{2point_momentum}). In particular $B$, expressed through $\lambda$ parameter associated with the size of propagating operator, gives a constant symmetric term in the Green's function, and $B_1,B_2$, expressed through $\lambda$ parameter related to the charge density, lead to an antisymmetric in time term in the Green's function.    
In double-scaling limit, the infinite series of moments for the partition and the two-point function diverge unless we restrict the chemical potential to be below a critical value. 
This contrasts with $\mu=0$ case, where all infinite series converge. 
This gave a range for the charge density of free fermions $Q_0^2\leq 1/(8\beta{\mathcal J})$, which is close to the estimate obtained by the direct large $p$ calculations in Section~\ref{section_largepSYK}. The two methods, direct large $p$ SYK calculations and double-scaled SYK, use different mathematical approaches, that both break at some finite chemical potential $\mu_{*}$. In the large $p$ calculations, a transcendental equation for a parameter $v$, which parametrized $\beta J$, does not have solution for a fixed $\mu$. In double-scaled SYK, the infinite sums for the moments of the Hamiltonian and the two-point function diverge at a finite $\mu$.
In both methods we need to adjust $\mu<\mu_{*}(\beta {\mathcal J})$ 
with increasing $\beta {\mathcal J}$. We found a self-consistent scheme in double-scaling SYK calculations by keeping the first order in all $\lambda$'s. The 
parameter $\lambda\to 0$ associated with the Hamiltonian size controls the semiclassical limit in double-scaled SYK, and generally it is analogous to ${\hbar}$ \cite{Double_scaled_SYK_small_lambda}.
This ensured complete agreement in the partition function and the two-point propagator between the two methods.     

Section~\ref{section_metric} presented the holographic mapping of the results of Section~\ref{section_largepSYK}. Here, we generalized earlier results in Refs.~\cite{Large_q,Large_q2,Large_q_Schwarzian,Double_scaled_SYK_small_lambda} for the Majorana SYK model. The holographic mapping is to a 
two-dimensional Jackiw-Teitelboim theory, but for the complex case we needed to introduce an additional $U(1)$ gauge field resulting in 2D Einstein-Maxwell-Dilaton theory \cite{Complex_SYK,Complex_SYK-Large_q, Sachdev:2019bjn,Complex_SYK_Density_of_states}. The metric is connected to the symmetric component 
of the fermion Green's function $g_s(\tau)$ ~\cite{Large_q,Large_q2,Large_q_Schwarzian,Double_scaled_SYK_small_lambda}. Combining the earlier results \cite{Complex_SYK,Complex_SYK-Large_q, Sachdev:2019bjn}, we connected the $U(1)$ gauge field to the antisymmetric component of the $U(1)$ gauge field in Eq.~(\ref{electric_field}). We also determined the modification of the profile of the dilaton of the Jackiw-Teitelboim theory from this electric field.
\begin{itemize}
\item
Generally Einstein-Maxwell-Dilaton model with $AdS_2$ background admits two types of dilaton solutions, those with running dilaton and those with constant dilaton. Both are asymptotically $AdS_2$, but constant dilaton solution only exist in the presence of the non zero electric charge and respects the full $AdS_2$ algebra. Such a
space is maximally symmetric and exhibits three Killing vectors: it is locally (and
asymptotically) $AdS_2$. The dilaton, which is not constant on $AdS_2$, explicitly breaks the
conformal symmetry: it breaks the $SL(2,R)$ algebra generated by
the Killing vectors to $U(1)$, giving rise to zero modes for reparametrizations with an action determined by Schwarzian. Taking the electric field to zero, we recovered the dilaton field of the Jackiw-Teitelboim theory which diverges near the boundary \cite{Large_q_Schwarzian} corresponding to the SYK Majorana fermions.  It is interesting that the solution for the dilaton field $e^{-\Phi}$ in Eq.~(\ref{hypergeometric}) and the eigenfuctions of the four fermion kernel and for the Casimir operator \cite{Large_q} are given by the hypergeometric functions 
${}_2F_1(\dots)$, showing a duality between the gravity and SYK theories. 
\item
Compared to the neutral case, in the complex SYK an effective action for the Green's function fluctuation in the kinematic space consists of a Liouville action plus an additional boundary action due to nonzero values of the Green's function and its derivative at the boundary. Mapping an effective action of Green's function fluctuations in the kinematic space Eq.~(\ref{effective_action3}) to the 2D Einstein-Maxwell-Dilaton theory Eq.~(\ref{2d_gravity}) we made the following identifications: a symmetric Green's function $g_s(\tau)$ with the metric, an antisymmetric Green's function $g_a(\tau)$ with an electric field, and total Green's function $g(\tau)=g_s(\tau)+g_a(\tau)$ with the dilaton field Eq.~(\ref{mapping3}). Then
Eq.~(\ref{Ricci_scalar}) for the scalar curvature reproduces Liouville equation of motion for the symmetric Green's function $g_s(\tau)$, and Eq.~(\ref{eom_constant3}) for the field strength gave the equation of motion for the antisymmetric Green's function $g_a(\tau)$. This proved holographic duality between the effective and gravity theories for equations of motion. The Ricci scalar is the Laplacian in the $AdS_2$ with radius $a=1/{\mathcal J}$ applied to the symmetric Green's function, that corresponds in the conformal limit SYK 
$\beta{\mathcal J}\gg1$ to the Casimir operator of $SL(2,R)$ group acting on the two times, which is the Laplacian in the $AdS_2$ space with radius $a=1$ \cite{Large_q}. This showed duality on the level of symmetry algebras also  between the complex double-scaled SYK and gravity theory \cite{Algebras}.
We found the same Ricci scalar as in the neutral case \cite{Double_scaled_SYK_small_lambda}. This happened due to the fact that the Liouville factor in the interaction $e^{g_s}$ and the metric included only thd symmetric Green's function; that leads to decoupling the $SL(2,R)$ and electric $U(1)$ degrees of freedom in the Schwarzian action \cite{Complex_SYK_Density_of_states}.
\item
Finally, we showed that a gravity dual of the complex double-scaled SYK in the regime $\lambda\to 0$ and $\beta{\mathcal J}$ is large but when ${1}/{\lambda\beta{\mathcal J}}\gg 1$ \cite{Algebras},
{\it i.e.\/} finite temperature/energy, 
is given by a semiclassical Jackiw-Teitelboim gravity theory with Maxwell term. It matched on to the low but finite temperature complex SYK in large $p$ limit.
The parameter $\lambda$ controls the semiclassical limit in the Liouville theory.
We performed perturbative calculations to linear order in $\lambda$ to obtain the same Green's functions in large $p$ and double-scaled complex SYK limits. In this approach,
the geometry is defined by a two-point function, the metric is given by the symmetric Green's function $g_s(\tau)$, producing two-dimensional Anti-de Sitter space-time at low $\beta{\mathcal J}\to \infty$ and finite temperatures. Since $g_s(0)$ does not contribute to this order in $\lambda$, the result for the Ricci curvature is the same for the neutral and charged cases. With increasing 
$\lambda$ fluctuations become large: it would be desirable to understand the geometry at finite $\lambda$ and the corresponding deformation of the Jackiw-Teitelboim gravity. 
\item As was noted in Ref.~\cite{Double_scaled_SYK_small_lambda}, there is an ambiguity in choosing anti-de Sitter or de Sitter signature for the metric. Here we chose anti-de Sitter geometry in the classical limit, that allowed us to identify the gravity dual as semiclassical JT gravity. Though in general de Sitter space should admit a similar holographic description, results on this front has been more limited. It is interesting to understand whether complex double-scaled SYK can be used as a model of low dimensional de Sitter holography.  
\end{itemize}
{\bf Note added} As we submitted our paper we became aware of two recent papers \cite{JiuciXu,YannicKruse} on the complex double-scaled SYK beyond semiclassical limit, exploring quantum regime of the model at finite 
$\lambda$.

\subsection*{Acknowledgements}
We thank M. Christos, M. Delgado, Y.Gu, M. Jariwala, S. Lee, L. Levitov, J. Maldacena, J. McGreevy, A. Mikhailov, N. Miller, E. Mishchenko, A. Patel, A. Ruckenstein, H. Verlinde for discussions; special thanks to Emil Akhmedov, Rashmish Mishra, Vladimir Narovlansky, Jan Zaanen and Konstantin Ziabliuk. E.G. thanks the Flatiron Institute for hospitality.
S. S. was supported  by the U.S. National Science Foundation grant No. DMR-2245246 and by the Simons Collaboration on Ultra-Quantum Matter which is a grant from the Simons Foundation (651440, S.S.).  G. T. was supported  by the U.S. Department of Energy Grant
No. DE-SC0010118.

\appendix

\section{Effective action for the complex SYK model}
\label{cSYKEffAct}

Starting with the original fermion path
integral for the Hamiltonian (\ref{complex_Hamiltonian}) and doing the Gaussian integral over the disorder one derives an effective action of the model, which is a bilocal action for
the fermions. Doing Hubbard-Stratonovich transformation by introducing a field $G$ and a
Lagrange multiplier field $\Sigma$ that sets $G(\tau_{1},\tau_{2})=\frac{1}{N}\sum_{j}\langle \textrm{T} \psi^{j}(\tau_1) \bar{\psi}_{j}(\tau_2)\rangle$ \cite{Complex_SYK,SY93}, one obtains 
the $G\Sigma$ partition function  
\begin{align}
\langle Z\rangle_{J} =\int DG D\Sigma {\rm e}^{-I[G,\Sigma]}\,,
\label{partition_function}
\end{align} 
where an effective $G\Sigma$ action for the complex SYK model is \cite{Complex_SYK_Density_of_states,Complex_SYK-Large_q}
\begin{align}
-I[G,\Sigma] =& N\log{\rm det}\left [\delta(\tau_1-\tau_2)(\partial_{\tau_1}-\mu)-\Sigma(\tau_1,\tau_2)\right]&& \nonumber\\
&-N \int d\tau_1 d\tau_2\left[ \Sigma(\tau_1,\tau_2)G(\tau_2,\tau_1)-\frac{J^2}{p} [G(\tau_1,\tau_2)]^{p/2}[G(\tau_2,\tau_1)]^{p/2} \right]\,.&&
\label{effective_action}
\end{align} 
In the large $N$ limit the effective action has a saddle point at the solutions $G, \Sigma$ of the Schwinger-Dyson equations
\begin{align}
&(\partial_{\tau_1}-\mu)G(\tau_1-\tau_2)+\int d\tau^{\prime}\Sigma(\tau_1-\tau^{\prime})G(\tau^{\prime}-\tau_2) = \delta(\tau_1-\tau_2), \notag\\
&\Sigma(\tau_1,\tau_2)=J^2 [G(\tau_1,\tau_2)]^{\frac{p}{2}}[G(\tau_2,\tau_1)]^{\frac{p}{2}-1}\,.
\label{eom}
 \end{align}
 In effective action Eq.~(\ref{effective_action}) $G$ and $\Sigma$ are fluctuating fields. 
The last term in Eq.~(\ref{effective_action}) can be written as
\begin{align}
N\beta\int_0^{\beta}d\tau\left[ \Sigma(\tau)G(\beta-\tau)+\Sigma(\beta-\tau)G(\tau)-\frac{J^2}{p} [G(\tau)]^{p/2}[G(\beta-\tau)]^{p/2} \right]\,,
 \end{align}
and the Schwinger-Dyson equations for the complex SYK model are  
\begin{align}
G(i\omega_{n})= \frac{1}{-i\omega_{n}-\mu-\Sigma(i\omega_{n})}, \quad \Sigma(\tau) = J^{2}G(\tau)^{\frac{p}{2}} G(\beta-\tau)^{\frac{p}{2}-1}\,,
\label{SD}
\end{align}
where $\omega_{n}=2\pi/\beta(n+1/2)$ are the Matsubara frequencies and 
\begin{align}
G(i\omega_{n}) = \int_{0}^{\beta}d\tau G(\tau)e^{ i\omega_{n} \tau}, \quad G(\tau)= \frac{1}{\beta}\sum_{n} G(i\omega_{n})e^{-i\omega_{n}\tau}\,.
\end{align}

Let us now discuss some general properties of the Green's function $G(\tau)$. We can represent it using the spectral density as 
\begin{align}
G(i\omega_{n})= \int_{-\infty}^{+\infty} \frac{d\Omega}{2\pi}\frac{\rho(\Omega)}{\Omega-i\omega_{n}}, \quad G(\tau) = \int_{-\infty}^{+\infty} \frac{d\Omega}{2\pi}\frac{\rho(\Omega)}{1+e^{-\beta \Omega}}e^{-\Omega \tau}\,.
\end{align}
From these equations we  find 
\begin{align}
&G(0^{+})+G(\beta^{-}) = \int_{-\infty}^{+\infty} \frac{d\Omega}{2\pi}  \rho(\Omega) \,, \quad  \partial_{\tau}G(0^{+})+\partial_{\tau}G(\beta^{-}) = - \int_{-\infty}^{+\infty} \frac{d\Omega}{2\pi} \Omega \rho(\Omega)\,.
\end{align}
On the other hand, for the large $i\omega_{n}$ we expect that 
\begin{align}
G(i\omega_{n}) \to \frac{1}{-i\omega_{n}} - \frac{\mu}{\omega_{n}^{2}}+\dots  = \frac{1}{-i\omega_{n}}\int_{-\infty}^{+\infty} \frac{d\Omega}{2\pi} \rho(\Omega)+
\frac{1}{\omega_{n}^{2}}\int_{-\infty}^{+\infty} \frac{d\Omega}{2\pi} \Omega \rho(\Omega) +\dots\,,
\end{align}
therefore we conclude that 
\begin{align}
G(0^{+})+G(\beta^{-}) = 1, \quad \partial_{\tau}G(0^{+})+\partial_{\tau}G(\beta^{-})  = \mu\,. 
\label{relmu}
\end{align}
Taking $J=0$ we obtain the resuls for the Green's function of free complex fermions  on the interval $0<\tau <\beta$:
\begin{align}
G_{0}(i\omega_{n})=\frac{1}{-i\omega_{n}-\mu}, \quad G_{0}(\tau)= \frac{e^{\mu\tau}}{1+e^{\beta \mu}}\,, 
\label{freeG}
\end{align}
and $Q_{0} = 1/2 - G_{0}(0^+)=\frac{1}{2}\tanh (\frac{\beta \mu}{2})$.


\section{Solving equation for the symmetric Green's function}
\label{SymGreenSolv}

We solve initial value problem Eq.~(\ref{symmetric}) for $g_s(\tau)$ 
\begin{align}
\partial_{\tau}^2g_s(\tau)=2\mathcal{J}^2e^{g_s(\tau)} \label{Eqgs1}
\end{align}
with the corresponding boundary conditions.
Multiplying both sides of the Eq.~(\ref{Eqgs1}) by $\partial_{\tau}g_s\equiv g_s^{\prime}$ 
\begin{align}
\frac{1}{2}\frac{d}{d\tau}(g_s^{\prime})^2=2\mathcal{J}^2e^{g_s}g_s^{\prime}
\label{solving_gs}
\end{align}
we can integrate Eq.~(\ref{solving_gs})
\begin{align}
g_s^{\prime}=2\mathcal{J}\sqrt{e^{g_s}+C}\,,
\end{align}
where we took positive sign for the square root. Integrating second time 
\begin{align}
\tanh^{-1}\left(\sqrt{1+\frac{e^{g_s}}{C}}\right)=
(-\mathcal{J}\tau+\tilde{C})\sqrt{C}\,,
\end{align}
where $C,\tilde{C}$ are constants of integration. Using trigonometric identity, rescaling 
$\mathcal{J}\sqrt{C}\to -C$ and introducing a new constant $\tau_0$ we have 
\begin{align}
e^{g_s(\tau)}=-\frac{C^2}{\mathcal{J}^2}\frac{1}{\cos^2(C(|\tau|+\tau_0))}\,.
\end{align}
Parameterizing this solution as in \cite{Large_q} and taking into account nonzero boundary conditions Eq.~(\ref{symmetric}) , we introduce
$C=\frac{\pi v}{\beta}$ with $\frac{C^2}{\mathcal{J}^2}=-\cos^2\left(\frac{\pi v}{2}\right)e^{4\pi v\tan\left(\frac{\pi v}{2}\right)}Q_0^2$ and $C\tau_0=-\frac{\pi v}{2}$, resulting in
\begin{align}
e^{g_s(\tau)}=\frac{\cos^2\left(\frac{\pi v}{2}\right)}{\cos^2\left(\frac{\pi v}{2}-\frac{\pi v |\tau|}{\beta}\right)}e^{4\pi v\tan\left(\frac{\pi v}{2}\right)Q_0^2}\,,
\end{align}
where 
\begin{align}
\frac{\pi v}{\cos\left(\frac{\pi v}{2}\right)}
e^{-2\pi v\tan\left(\frac{\pi v}{2}\right)Q_0^2}= \beta\mathcal{J}\,,
\end{align}
which is a solution for the symmetric part $g_s(\tau)$ given in Eq.~(\ref{g_solution},\ref{betaJ_solution}).

\section{Comparison between the large \texorpdfstring{$p$}{p} and exact Green's functions}
\label{GexactGp}
In this Appendix we compare the large $p$ approximation $G_{\textrm{large}\,p}(\tau)$ in Eq.~(\ref{largepmu})  with the exact large $N$ complex SYK Green's function $G_{\textrm{exact}}(\tau)$.
We obtain $G_{\textrm{exact}}(\tau)$ by numerically solving the large $N$ Schwinger-Dyson equations in (\ref{SD}).
We solve this system using an iterative procedure, first introduced in Appendix G of \cite{Large_q}, which, given an approximate solution $(G^{(j)},\Sigma^{(j)})$ to Eq.~(\ref{SD}), 
generates another one via a weighted update  
\begin{align}
G^{(j+1)}(i\omega_n)= w\frac{1}{-i\omega_{n}-\mu-\Sigma^{(j)}(i\omega_{n})} + (1-w)G^{(j)}(i\omega_n)\,,
\label{numerics}
\end{align}
where $w$ is the weighting parameter. 

\begin{figure}[h!]
\includegraphics[width=1.0\textwidth]{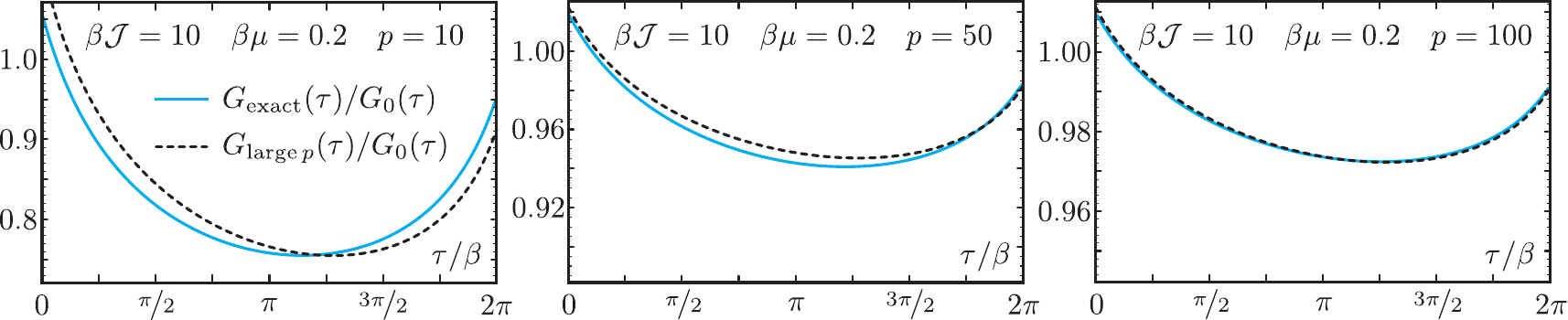}
\caption{Comparison of the exact complex SYK Green’s function (blue solid lines; obtained by numerically solving the Schwinger–Dyson equations (\ref{SD})) with the large $p$ approximation Eq.~(\ref{largepmu}) (black dashed lines)  at $\beta\mathcal{J}=10$ and $\beta\mu=0.2$. Panels correspond to $p=10$,  $p=50$, and $p=100$. All curves are normalized to $G_{0}(\tau)$.}
\label{figGexGp}
\end{figure}

In Figure \ref{figGexGp}, we present the numerical solution for the complex SYK Green's function together with its large $p$ approximation Eq.~(\ref{largepmu}) for $\beta \mathcal{J} =10$ and $\beta\mu = 0.2 $ with $p=10$, $p=50$ and $p=100$. For clarity, both functions are normalized by the free fermion Green's function $G_{0}(\tau)$ in Eq.~(\ref{freeG}). 
We notice that a good agreement between the large $p$ and the exact Green's functions is achieved only for sufficiently large values of $p$. This is expected, since expanding the asymmetry parameter $\mathscr{E}$ in Eq.~(\ref{Elp1}) at large $\beta \mathcal{J}$ using Eq.~(\ref{largeJmuexp}) gives
\begin{align}
2\pi \mathscr{E} = \beta \mu - \frac{1}{p} \Big(\frac{8 \beta \mathcal{J}}{  v_{1}} + O(1/\beta \mathcal{J})\Big)Q_{0} + O(1/p^2)\,. \label{Elp2}
\end{align}
For the large $p$ approximation to be reliable, the $1/p$ term must be much smaller than the leading term $\beta \mu$.  Using that $Q_{0}\approx \beta \mu /4$ for small $\beta \mu$ we obtain the condition for $p$
\begin{align}
p \gg \beta \mathcal{J} e^{2\kappa_{0}}\,,
\end{align}
where  $v_{1} \approx 2 e^{-2\kappa_{0}}$ and  $Q_{0}^{2} = \kappa_{0}/(\beta \mathcal{J})$ with  $\kappa_{0}\leq 0.184$. A similar behavior was observed for the eigenvalues of the complex SYK kernel in the symmetric sector in the large $p$ limit 
\cite{Complex_SYK_Density_of_states}.
In contrast, this behavior is absent in the Majorana SYK  (or antisymmetric sector of the complex SYK) model and the large $p$ approximation works well even for $p=4$ \cite{Large_q,Complex_SYK_Density_of_states,PhysRevD.99.026010}.

\section{Phase transition}
\label{AppPhTr}
In this Appendix we find numerical solutions of the Green's function for the complex SYK at different chemical potentials and various temperatures. We use an iterative procedure introduced in \cite{Large_q}. We start iteration with the conformal solution $G_0=G_c$.
 \begin{figure}[h!]
    \centering
\includegraphics[width=0.45\textwidth]{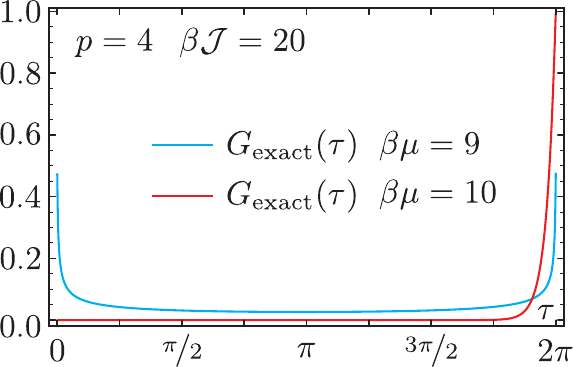} 
\qquad
\includegraphics[width=0.45\textwidth]{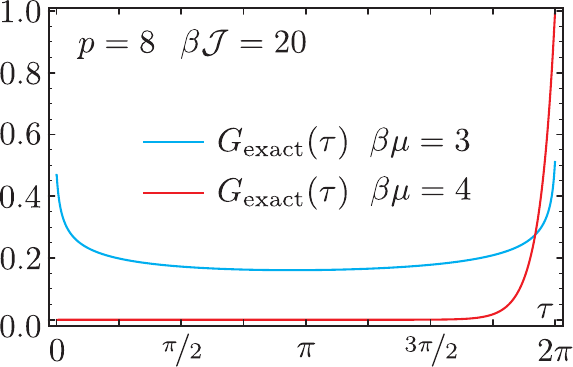} 
\caption{Left: Numerical solution for the Green’s function for $p = 4$ and $\beta \mathcal{J} = 20$. The phase transition occurs between $\beta \mu = 9$ and $\beta \mu = 10$. Right: Numerical solution for the Green’s function for $p = 8$ and $\beta \mathcal{J} = 20$. The phase transition occurs between $\beta \mu = 3$ and $\beta \mu = 4$.}
 \label{Fig1}
 \end{figure}
In Figure \ref{Fig1} we display numerical solution for the Green's function at the finite temperature for periodic times $0<\tau<\beta$. 
We distinguish two regimes:  
the SYK-like regime with high entropy at small chemical potential $\mu$ and the harmonic oscillator-like regime with low entropy at large $\mu$. 
In the SYK-like regime numerical solution is close to the conformal Green's function, which is valid
for small temperatures and strong coupling $\beta J\gg 1$. 
In particular they almost coincide away from the end points $\tau=0,\beta$. 
This confirms our analytical finding in Section~\ref{LargepandGc} that in the limit of large $\beta J$ the Green's function coincides with the conformal solution to the leading order. 

At larger $\mu$ there is a first order phase transition to an oscillator-like solution 
which decays exponentially with a characteristic energy gap in the exponent $\sim{\rm e}^{-\Delta\tau}$. The oscillator solution is trivial and corresponds to a low entropy state.    
There is a line of first-order phase transitions terminating at a critical point for a positive temperature.
The same two regimes and the phase transition between them were obtained numerically for the complex SYK in \cite{Complex_SYK_Phase_transition,Complex_SYK_Phase_transition2}.

\section{Luttinger relation in the large $p$ limit}
\label{GpLuttRel}
The charge $Q$ and asymmetry parameter $\mathscr{E}$ in the conformal approximation of the complex SYK Green's function are related through the Luttinger relation \cite{GPS, Complex_SYK_Density_of_states, Tikhanovskaya_2021}:
\begin{align}
Q = \frac{1}{2\pi i} \ln \left(\frac{\cos(\frac{\pi}{p} - i \pi \mathscr{E})}{\cos(\frac{\pi}{p} + i \pi \mathscr{E})}\right) + \frac{1-\frac{2}{p}}{4i}\Big(\tan (\frac{\pi}{p}+i\pi \mathscr{E})-\tan(\frac{\pi}{p}-i\pi \mathscr{E})\Big)\,, \label{LutRel}
\end{align}
and in the large $p$ limit this relation reads 
\begin{align}
Q =  \frac{1}{2}\tanh(\pi  \mathscr{E}) + O(1/p^2)\,. \label{relQE}
\end{align}
While it is possible to relate $Q$ to $\mathscr{E}$ as in Eq.~(\ref{LutRel}), there is no analytical result for the relation between $Q$ and $\mu$, or between $\mathscr{E}$ and $\mu$. In the large $p$ limit, the relation  between  $Q$ and $Q_{0}$ is obtained from the large $p$ solution Eq.~(\ref{largepmu}). Using that $G(0^{+})=1/2-Q$ and $G_{0}(0^+) = 1/2-Q_{0}$, we find  
\begin{align}
Q = Q_{0} - \frac{1}{p} 4\pi v \tan\Big(\frac{\pi v}{2}\Big) Q_{0}\Big(\frac{1}{4}-Q_{0}^2\Big) + O(1/p^2)\,. \label{Qlp2}
\end{align}
One can verify that Eq.~(\ref{Qlp2}) and Eq.~(\ref{Elp1}) indeed satisfy the Luttinger relation Eq.~(\ref{relQE}) in the large $p$ limit.

\section{Geodesic and geodesic deviation equations}
\label{AppGeodEq}

We derive equations of geodesic lines for the metric
\begin{align}
ds^2=e^{g_s}\left(-c^2dt^2+dz^2\right)\,,
\label{metric1}
\end{align}
where we insert speed of light $c$, that means 
$G_{\mu\nu}=(c^2e^{g_s},e^{g_s})$ and $G^{\mu\nu}=(-e^{-g_s}/c^2,e^{-g_s})$, and
$g_s(z)$ depends only on $z$ coordinate as a solution of equation of motion 
\begin{align}
e^{g_s(z)}=\left(\frac{\pi v}{\beta}\right)^2
\frac{1}{\mathcal{J}^2\cos^2\left(\frac{\pi v}{2}-\frac{\pi v z}{\beta}\right)}\,.
\label{}
\end{align}
Christoffel symbols
$\Gamma^{k}_{ij}=\frac{1}{2}g^{kl}\left(\partial_ig_{jl}+\partial_{j}g_{il}-\partial_{l}g_{ij}\right)$, are given for the metric Eq.~(\ref{metric1})
\begin{align}
\Gamma^{z}_{tt}=\frac{1}{2}c^2\partial_zg_s\,,\quad 
\Gamma^{z}_{zz}=\Gamma^{t}_{zt}=\Gamma^{t}_{tz}=\frac{1}{2}\partial_zg_s.
\label{}
\end{align}
Geodesic equation is given $\frac{d^2x^{\lambda}}{dq^2}+\Gamma^{\lambda}_{\mu\nu}\frac{dx^{\mu}}{dq}\frac{dx^{\nu}}{dq}=0$, where variable $q$ parametrizes the particle world-line. Using Christoffel symbols geodesic equations are
\begin{align}
 & \frac{d^2t}{dq^2}+\partial_zg_s\frac{dz}{dq}\frac{dt}{dq}=0 \label{geodesic_time},\\
 & \frac{d^2z}{dq^2}+\frac{1}{2}\partial_zg_s\left(\frac{dz}{dq}\right)^2+\frac{1}{2}c^2\partial_zg_s\left(\frac{dt}{dq}\right)^2=0
 \label{geodesic_space}.
\end{align}
To solve equation for the time variable (\ref{geodesic_time}) we divide it by $\frac{dt}{dq}$ 
\begin{align}
\frac{d}{dq}\left[\ln \frac{dt}{dq}+g_s\right]=0, 
\end{align}
that gives constant of motion $E$, energy of a test particle,
\begin{align}
\frac{E}{mc^2}=e^{g_s}\frac{dt}{d\tau}, 
\label{energy}
\end{align}
where $\tau$ is a proper time introduced instead of $q$ variable, and we reinstated $c$ for the dimension. We can write the metric Eq.~(\ref{metric1}) for the proper time interval
\begin{align}
c^2d\tau^2=e^{g_s}(-c^2dt^2+dz^2),
\end{align}
substituting the energy Eq.~(\ref{energy}), it gives
\begin{align}
\left(\frac{dz}{d\tau}\right)^2=c^2\left(\frac{E}{mc^2}e^{-g_s}\right)^2
+c^2e^{-g_s}\,.
\end{align}
Formal solution to the proper time is
\begin{align}
\tau=\int \frac{dz}{\pm c\frac{E}{mc^2}e^{-g_s}\sqrt{1+\left(\frac{mc^2}{E}\right)^2e^{g_s}}}\,.
\end{align}
Using Eq.~(\ref{energy}) $dt=\frac{E}{mc^2}e^{-g_s}d\tau$, we have for the time of an observer
\begin{align}
t=\int \frac{dz}{\pm c\sqrt{1+\left(\frac{mc^2}{E}\right)^2e^{g_s}}}
\end{align}
and
\begin{align}
\frac{dz}{cdt}=\pm\sqrt{1+\left(\frac{mc^2}{E}\right)^2e^{g_s}}\,.
\end{align}
For photonic geodesic $m=0$, it takes finite observer's time to reach any point in space, meaning that there is no horizon. Light propagates along trajectories $dz=\pm cdt$ and there is no black hole in the metric solution Eq.~(\ref{metric1}).

Next we analyze the geodesic deviation equations. Riemann tensor
$R^{\alpha}_{\beta\mu\nu}=\partial_{\mu}\Gamma^{\alpha}_{\beta\nu}
-\partial_{\nu}\Gamma^{\alpha}_{\beta\mu}+\Gamma^{\alpha}_{\sigma\mu}
\Gamma^{\sigma}_{\beta\nu}-\Gamma^{\alpha}_{\sigma\nu}\Gamma^{\sigma}_{\beta\mu}$, are given for the metric Eq.~(\ref{metric1})
\begin{align}
R^{t}_{ztz}=-\frac{1}{2}\partial^2_z g_s,\, \quad
R^{z}_{tzt}=\frac{1}{2}c^2\partial^2_z g_s\,,
\end{align}
One contracts the first and third indices in Riemann tensor to get the Ricci curvature tensor $R_{ik}=R^{l}_{ilk}$,
\begin{align}
R_{zz}=-\frac{1}{2}\partial^2_z g_s,\,\quad
R_{tt}=\frac{1}{2}c^2\partial^2_z g_s\,.
\end{align}
Consider a relative acceleration of two test particles located at infinitely close geodesic lines $\Gamma(\nu)$ and $\Gamma(\nu+d\nu)$ with coordinates $x^i(\tau,\nu)$ where $\tau$ is affine parameter along the geodesic line and $\nu$ is perpendicular to a tangent vector and remains constant along the line. We describe a distance between these two geodesic lines by an infinitesimal deviation vector $\eta^i$ satisfying an equation \cite{Misner_Thorne_Wheeler}
\begin{align}
\frac{D^2\eta^i}{d\tau^2}+R^{i}_{jkm}u^j\eta^k u^m=0\,,
\end{align}
where $R^i_{jkm}$ is Riemann tensor, $u^i=\left(\frac{dt}{d\tau},\frac{dz}{d\tau},0,0\right)$ is a velocity of one particle relative to the other. Geodesic deviation vector $\eta^i$ is proportional to tidal forces, corresponding to stretching when it is positive  and compression when it is negative. Deviation of geodesic lines is given
\begin{align}
\frac{D^2\eta^t}{d\tau^2}&=\frac{1}{2}\partial^2_z g_s u^z(u^t+u^z)>0,\\
\frac{D^2\eta^z}{d\tau^2}&=-\frac{1}{2}c^2\partial^2_z g_s u^z(u^t+u^z)<0,
\end{align}
where components of velocity are positive and acceleration difference $D^2\eta^z/d\tau^2$ is proportional to
\begin{align}
\partial^2_z g_s=\frac{2\left(\frac{\pi v}{\beta}\right)^2}{\cos^2\left(\frac{\pi v}{2}-\frac{\pi v z}{\beta}\right)}\,,
\end{align}
which has singularities at $\frac{\pi v}{2}-\frac{\pi v z}{\beta}=\frac{\pi}{2}+\pi n,\,n\in \mathbb{Z}$. This means that a force acting on different parts of an object separated by $\eta^z$ increases, stress grows leading that an object being torn apart when approaching singularity. It is a physical singularity which can not be removed by a transformation. It is a "naked" singularity which does not have a  horizon, therefore matter can escape.

\bibliography{dss.bib}

@article{Complex_SYK,
    author = "Sachdev, Subir",
    title = "{Bekenstein-Hawking Entropy and Strange Metals}",
    eprint = "1506.05111",
    archivePrefix = "arXiv",
    primaryClass = "hep-th",
    doi = "10.1103/PhysRevX.5.041025",
    journal = "Phys. Rev. X",
    volume = "5",
    number = "4",
    pages = "041025",
    year = "2015"
}

@article{SS10,
      author         = "Sachdev, Subir",
      title          = "{Holographic metals and the fractionalized Fermi liquid}",
      journal        = "Phys. Rev. Lett.",
      volume         = "105",
      pages          = "151602",
      doi            = "10.1103/PhysRevLett.105.151602",
      year           = "2010",
      eprint         = "1006.3794",
      archivePrefix  = "arXiv",
      primaryClass   = "hep-th",
      SLACcitation   = "%%CITATION = ARXIV:1006.3794;%%",
}

@article{SYK_Review2,
    author = "Rosenhaus, Vladimir",
    title = "{An introduction to the SYK model}",
    eprint = "1807.03334",
    archivePrefix = "arXiv",
    primaryClass = "hep-th",
    doi = "10.1088/1751-8121/ab2ce1",
    journal = "J. Phys. A",
    volume = "52",
    pages = "323001",
    year = "2019"
}

@article{DSSYK_Review,
    author = "Berkooz, Micha and Mamroud, Ohad",
    title = "{A Cordial Introduction to Double Scaled SYK}",
    journal = {},
    eprint = "2407.09396",
    archivePrefix = "arXiv",
    primaryClass = "hep-th",
    month = "7",
    year = "2024"
}

@ARTICLE{SY93,
       author = {{Sachdev}, Subir and {Ye}, Jinwu},
        title = "{Gapless spin-fluid ground state in a random quantum Heisenberg magnet}",
      journal = {Phys. Rev. Lett.},
         year = 1993,
        month = may,
       volume = {70},
       number = {21},
        pages = {3339-3342},
          doi = {10.1103/PhysRevLett.70.3339},
archivePrefix = {arXiv},
       eprint = {cond-mat/9212030},
 primaryClass = {cond-mat},
       adsurl = {https://ui.adsabs.harvard.edu/abs/1993PhRvL..70.3339S}
}

@article{Large_q,
    author = "Maldacena, Juan and Stanford, Douglas",
    title = "{Remarks on the Sachdev-Ye-Kitaev model}",
    eprint = "1604.07818",
    archivePrefix = "arXiv",
    primaryClass = "hep-th",
    doi = "10.1103/PhysRevD.94.106002",
    journal = "Phys. Rev. D",
    volume = "94",
    number = "10",
    pages = "106002",
    year = "2016"
}

@article{Large_q_Schwarzian,
    author = "Maldacena, Juan and Stanford, Douglas and Yang, Zhenbin",
    title = "{Conformal symmetry and its breaking in two dimensional Nearly Anti-de-Sitter space}",
    eprint = "1606.01857",
    archivePrefix = "arXiv",
    primaryClass = "hep-th",
    doi = "10.1093/ptep/ptw124",
    journal = "PTEP",
    volume = "2016",
    number = "12",
    pages = "12C104",
    year = "2016"
}

@article{Double_scaled_SYK,
    author = "Berkooz, Micha and Isachenkov, Mikhail and Narovlansky, Vladimir and Torrents, Genis",
    title = "{Towards a full solution of the large N double-scaled SYK model}",
    eprint = "1811.02584",
    archivePrefix = "arXiv",
    primaryClass = "hep-th",
    doi = "10.1007/JHEP03(2019)079",
    journal = "JHEP",
    volume = "03",
    pages = "079",
    year = "2019"
}

@article{DSSYK2,
    author = "Berkooz, Micha and Narayan, Prithvi and Simon, Joan",
    title = "{Chord diagrams, exact correlators in spin glasses and black hole bulk reconstruction}",
    eprint = "1806.04380",
    archivePrefix = "arXiv",
    primaryClass = "hep-th",
    doi = "10.1007/JHEP08(2018)192",
    journal = "JHEP",
    volume = "08",
    pages = "192",
    year = "2018"
}

@article{Double_scaled_complex_SYK,
    author = "Berkooz, Micha and Narovlansky, Vladimir and Raj, Himanshu",
    title = "{Complex Sachdev-Ye-Kitaev model in the double scaling limit}",
    eprint = "2006.13983",
    archivePrefix = "arXiv",
    primaryClass = "hep-th",
    doi = "10.1007/JHEP02(2021)113",
    journal = "JHEP",
    volume = "02",
    pages = "113",
    year = "2021"
}

@article{Double_scaled_SYK_small_lambda,
    author = "Goel, Akash and Narovlansky, Vladimir and Verlinde, Herman",
    title = "{Semiclassical geometry in double-scaled SYK}",
    eprint = "2301.05732",
    archivePrefix = "arXiv",
    primaryClass = "hep-th",
    doi = "10.1007/JHEP11(2023)093",
    journal = "JHEP",
    volume = "11",
    pages = "093",
    year = "2023"
}

@article{Double_scaled_SYK_small_lambda2,
    author = "Mukhametzhanov, Baur",
    title = "{Large p SYK from chord diagrams}",
    eprint = "2303.03474",
    archivePrefix = "arXiv",
    primaryClass = "hep-th",
    doi = "10.1007/JHEP09(2023)154",
    journal = "JHEP",
    volume = "09",
    pages = "154",
    year = "2023"
}

@article{Double_scaled_SYK_small_lambda3,
    author = "Okuyama, Kazumi and Suzuki, Kenta",
    title = "{Correlators of double scaled SYK at one-loop}",
    eprint = "2303.07552",
    archivePrefix = "arXiv",
    primaryClass = "hep-th",
    reportNumber = "RUP-23-5",
    doi = "10.1007/JHEP05(2023)117",
    journal = "JHEP",
    volume = "05",
    pages = "117",
    year = "2023"
}

@article{Complex_SYK_Density_of_states,
    author = "Gu, Yingfei and Kitaev, Alexei and Sachdev, Subir and Tarnopolsky, Grigory",
    title = "{Notes on the complex Sachdev-Ye-Kitaev model}",
    eprint = "1910.14099",
    archivePrefix = "arXiv",
    primaryClass = "hep-th",
    doi = "10.1007/JHEP02(2020)157",
    journal = "JHEP",
    volume = "02",
    pages = "157",
    year = "2020"
}

@article{Large_q2,
    author = "Das, Sumit R. and Ghosh, Animik and Jevicki, Antal and Suzuki, Kenta",
    title = "{Three Dimensional View of Arbitrary $q$ SYK models}",
    eprint = "1711.09839",
    archivePrefix = "arXiv",
    primaryClass = "hep-th",
    doi = "10.1007/JHEP02(2018)162",
    journal = "JHEP",
    volume = "02",
    pages = "162",
    year = "2018"
}

@article{Large_q_Liouville,
    author = "Cotler, Jordan S. and Gur-Ari, Guy and Hanada, Masanori and Polchinski, Joseph and Saad, Phil and Shenker, Stephen H. and Stanford, Douglas and Streicher, Alexandre and Tezuka, Masaki",
    title = "{Black Holes and Random Matrices}",
    eprint = "1611.04650",
    archivePrefix = "arXiv",
    primaryClass = "hep-th",
    reportNumber = "SU-ITP-16-19, SU-ITP-16/19, YITP-16-124",
    doi = "10.1007/JHEP05(2017)118",
    journal = "JHEP",
    volume = "05",
    pages = "118",
    year = "2017",
    note = "[Erratum: JHEP 09, 002 (2018)]"
}

@article{Kinematic_space,
    author = "Czech, Bart\l{}omiej and Lamprou, Lampros",
    title = "{Holographic definition of points and distances}",
    eprint = "1409.4473",
    archivePrefix = "arXiv",
    primaryClass = "hep-th",
    doi = "10.1103/PhysRevD.90.106005",
    journal = "Phys. Rev. D",
    volume = "90",
    pages = "106005",
    year = "2014"
}

@article{Kinematic_space2,
    author = "Czech, Bartlomiej and Lamprou, Lampros and McCandlish, Samuel and Sully, James",
    title = "{Integral Geometry and Holography}",
    eprint = "1505.05515",
    archivePrefix = "arXiv",
    primaryClass = "hep-th",
    reportNumber = "SU-ITP-15-07",
    doi = "10.1007/JHEP10(2015)175",
    journal = "JHEP",
    volume = "10",
    pages = "175",
    year = "2015"
}

@article{Kinematic_space3,
    author = "Czech, Bartlomiej and Lamprou, Lampros and McCandlish, Samuel and Mosk, Benjamin and Sully, James",
    title = "{A Stereoscopic Look into the Bulk}",
    eprint = "1604.03110",
    archivePrefix = "arXiv",
    primaryClass = "hep-th",
    reportNumber = "SU-ITP-16-07",
    doi = "10.1007/JHEP07(2016)129",
    journal = "JHEP",
    volume = "07",
    pages = "129",
    year = "2016"
}

@article{2d_Maxwell_Dilaton,
    author = "Castro, Alejandra and Grumiller, Daniel and Larsen, Finn and McNees, Robert",
    title = "{Holographic Description of AdS(2) Black Holes}",
    eprint = "0809.4264",
    archivePrefix = "arXiv",
    primaryClass = "hep-th",
    reportNumber = "MIT-CTP-3981",
    doi = "10.1088/1126-6708/2008/11/052",
    journal = "JHEP",
    volume = "11",
    pages = "052",
    year = "2008"
}

@article{2dGravity,
    author = "Grumiller, D. and Kummer, W. and Vassilevich, D. V.",
    title = "{Dilaton gravity in two-dimensions}",
    eprint = "hep-th/0204253",
    archivePrefix = "arXiv",
    reportNumber = "TUW-02-01",
    doi = "10.1016/S0370-1573(02)00267-3",
    journal = "Phys. Rept.",
    volume = "369",
    pages = "327--430",
    year = "2002"
}

@article{2dGravity2,
    author = "Grumiller, Daniel and Meyer, Rene",
    title = "{Ramifications of lineland}",
    eprint = "hep-th/0604049",
    archivePrefix = "arXiv",
    reportNumber = "LU-ITP-2006-004",
    journal = "Turk. J. Phys.",
    volume = "30",
    pages = "349--378",
    year = "2006"
}

@article{Maxwell_Dilaton0,
    author = "Hartman, Thomas and Strominger, Andrew",
    title = "{Central Charge for AdS(2) Quantum Gravity}",
    eprint = "0803.3621",
    archivePrefix = "arXiv",
    primaryClass = "hep-th",
    doi = "10.1088/1126-6708/2009/04/026",
    journal = "JHEP",
    volume = "04",
    pages = "026",
    year = "2009"
}

@article{CGPS,
    author = "Chowdhury, Debanjan and Georges, Antoine and Parcollet, Olivier and Sachdev, Subir",
    title = "{Sachdev-Ye-Kitaev models and beyond: Window into non-Fermi liquids}",
    eprint = "2109.05037",
    archivePrefix = "arXiv",
    primaryClass = "cond-mat.str-el",
    doi = "10.1103/RevModPhys.94.035004",
    journal = "Rev. Mod. Phys.",
    volume = "94",
    number = "3",
    pages = "035004",
    year = "2022"
}

@article{Kitaev_Suh,
    author = "Kitaev, Alexei and Suh, S. Josephine",
    title = "{The soft mode in the Sachdev-Ye-Kitaev model and its gravity dual}",
    eprint = "1711.08467",
    archivePrefix = "arXiv",
    primaryClass = "hep-th",
    doi = "10.1007/JHEP05(2018)183",
    journal = "JHEP",
    volume = "05",
    pages = "183",
    year = "2018"
}

@ARTICLE{GPS,
       author = {{Georges}, A. and {Parcollet}, O. and {Sachdev}, S.},
        title = "{Quantum fluctuations of a nearly critical Heisenberg spin glass}",
      journal = {Phys. Rev. B},
         year = 2001,
        month = apr,
       volume = {63},
       number = {13},
          eid = {134406},
        pages = {134406},
          doi = {10.1103/PhysRevB.63.134406},
archivePrefix = {arXiv},
       eprint = {cond-mat/0009388},
 primaryClass = {cond-mat.str-el},
       adsurl = {https://ui.adsabs.harvard.edu/abs/2001PhRvB..63m4406G}
}

@article{Algebras,
    author = "Lin, Henry W. and Stanford, Douglas",
    title = "{A symmetry algebra in double-scaled SYK}",
    eprint = "2307.15725",
    archivePrefix = "arXiv",
    primaryClass = "hep-th",
    doi = "10.21468/SciPostPhys.15.6.234",
    journal = "SciPost Phys.",
    volume = "15",
    number = "6",
    pages = "234",
    year = "2023"
}

@book{Misner_Thorne_Wheeler,
  title        = {Gravitation},
  author       = {Misner, Charles W. and Thorne, Kip S. and Wheeler, John Archibald},
  year         = {1973},
  publisher    = {W. H. Freeman},
  address      = {San Francisco},
  isbn         = {9780716703440},
  note         = {ISBN-13: 978-0716703440}
}

@article{Complex_SYK_Phase_transition,
    author = "Ferrari, Frank and Schaposnik Massolo, Fidel I.",
    title = "{Phases Of Melonic Quantum Mechanics}",
    eprint = "1903.06633",
    archivePrefix = "arXiv",
    primaryClass = "hep-th",
    doi = "10.1103/PhysRevD.100.026007",
    journal = "Phys. Rev. D",
    volume = "100",
    number = "2",
    pages = "026007",
    year = "2019"
}

@article{Complex_SYK_Phase_transition2,
    author = "Azeyanagi, Tatsuo and Ferrari, Frank and Schaposnik Massolo, Fidel I.",
    title = "{Phase Diagram of Planar Matrix Quantum Mechanics, Tensor, and Sachdev-Ye-Kitaev Models}",
    eprint = "1707.03431",
    archivePrefix = "arXiv",
    primaryClass = "hep-th",
    doi = "10.1103/PhysRevLett.120.061602",
    journal = "Phys. Rev. Lett.",
    volume = "120",
    number = "6",
    pages = "061602",
    year = "2018"
}

@article{Complex_SYK-Large_q,
    author = "Davison, Richard A. and Fu, Wenbo and Georges, Antoine and Gu, Yingfei and Jensen, Kristan and Sachdev, Subir",
    title = "{Thermoelectric transport in disordered metals without quasiparticles: The Sachdev-Ye-Kitaev models and holography}",
    eprint = "1612.00849",
    archivePrefix = "arXiv",
    primaryClass = "cond-mat.str-el",
    doi = "10.1103/PhysRevB.95.155131",
    journal = "Phys. Rev. B",
    volume = "95",
    number = "15",
    pages = "155131",
    year = "2017"
}

@article{Sachdev:2019bjn,
    author = "Sachdev, Subir",
    title = "{Universal low temperature theory of charged black holes with AdS$_2$ horizons}",
    eprint = "1902.04078",
    archivePrefix = "arXiv",
    primaryClass = "hep-th",
    doi = "10.1063/1.5092726",
    journal = "J. Math. Phys.",
    volume = "60",
    number = "5",
    pages = "052303",
    year = "2019"
}

@article{2SYK_Large_q,
    author = "Maldacena, Juan and Qi, Xiao-Liang",
    title = "{Eternal traversable wormhole}",
    journal = {},
    eprint = "1804.00491",
    archivePrefix = "arXiv",
    primaryClass = "hep-th",
    month = "4",
    year = "2018"
}

@article{2SYK_Large_q3,
    author = "Maldacena, Juan and Milekhin, Alexey",
    title = "{SYK wormhole formation in real time}",
    eprint = "1912.03276",
    archivePrefix = "arXiv",
    primaryClass = "hep-th",
    doi = "10.1007/JHEP04(2021)258",
    journal = "JHEP",
    volume = "04",
    pages = "258",
    year = "2021"
}

@article{SYK_Review,
	author = {{Kitaev}, A. Y.},
	title = "{Talks at KITP, University of California, Santa Barbara}", 
	journal = {Entanglement in Strongly-Correlated Quantum Matter},
	year = {2015},
	url = {http://online.kitp.ucsb.edu/online/entangled15/}
	}

@article{PhysRevD.99.026010,
  title = {Large $q$ expansion in the Sachdev-Ye-Kitaev model},
  author = {Tarnopolsky, Grigory},
  journal = {Phys. Rev. D},
  volume = {99},
  issue = {2},
  pages = {026010},
  numpages = {5},
  year = {2019},
  month = {Jan},
  publisher = {American Physical Society},
  doi = {10.1103/PhysRevD.99.026010},
  url = {https://link.aps.org/doi/10.1103/PhysRevD.99.026010}
}

@article{PhysRevLett.120.061602,
  title = {Phase Diagram of Planar Matrix Quantum Mechanics, Tensor, and Sachdev-Ye-Kitaev Models},
  author = {Azeyanagi, Tatsuo and Ferrari, Frank and Massolo, Fidel I. Schaposnik},
  journal = {Phys. Rev. Lett.},
  volume = {120},
  issue = {6},
  pages = {061602},
  numpages = {6},
  year = {2018},
  month = {Feb},
  publisher = {American Physical Society},
  doi = {10.1103/PhysRevLett.120.061602},
  url = {https://link.aps.org/doi/10.1103/PhysRevLett.120.061602}
}

@article{Tikhanovskaya_2021,
   title={Excitation spectra of quantum matter without quasiparticles. I. Sachdev-Ye-Kitaev models},
   volume={103},
   ISSN={2469-9969},
   url={http://dx.doi.org/10.1103/PhysRevB.103.075141},
   DOI={10.1103/physrevb.103.075141},
   number={7},
   journal={Physical Review B},
   publisher={American Physical Society (APS)},
   author={Tikhanovskaya, Maria and Guo, Haoyu and Sachdev, Subir and Tarnopolsky, Grigory},
   year={2021},
   month=feb }

@article{JiuciXu,
    author = "van der Heijden, Jeremy and Verlinde, Erik and Xu, Jiuci",
    title = "{Quantum Symmetry and Geometry in Double-Scaled SYK}",
    eprint = "2511.08743",
    archivePrefix = "arXiv",
    primaryClass = "hep-th",
    month = "11",
    year = "2025"
}

@article{YannicKruse,
    author = "Forste, Stefan and Kruse, Yannic and Natu, Saurabh",
    title = "{Grand Canonical vs Canonical Krylov Complexity in Double-Scaled Complex SYK Model}",
    eprint = "2512.07715",
    archivePrefix = "arXiv",
    primaryClass = "hep-th",
    reportNumber = "BONN--TH--2025--35",
    month = "12",
    year = "2025"
}

@mastersthesis{Arundine:2025mcu,
    author = "Arundine, Mattia",
    title = "{The DSSYK Model: Charge and Holography}",
    eprint = "2512.21366",
    archivePrefix = "arXiv",
    primaryClass = "hep-th",
    type = "Master's Thesis",
    month = "12",
    year = "2025"
}

\end{document}